\documentclass[aps,preprint,epsfig,rotate]{revtex4}

\begin{document}
%
\title{General principles of Hamiltonian formulations of the metric gravity}
 \author{Alexei M. Frolov}
 \email[E--mail address: ]{alex1975frol@gmailcom}

\affiliation{Department of Applied Mathematics, \\
 University of Western Ontario, London, Ontario N6H 5B7, Canada}

\date{\today}

\begin{abstract}

Fundamental principles of some successful Hamiltonian approaches, which were developed to describe free 
gravitational field(s) in the metric gravity, are formulated and discussed. By using the standard $\Gamma 
- \Gamma$ Lagrangian ${\cal L}_{\Gamma - \Gamma}$ of the metric GR we properly introduce all momenta of 
the metric gravitational field and derive the both canonical $H_C$ and total $H_t$ Hamiltonians of the 
metric GR. We also developed an effective method which is used to determine various Poisson brackets 
between analytical functions of the basic dynamical variables, i.e., generalized coordinates 
$g_{\alpha\beta}$ and momenta $\pi^{\mu\nu}$. In general, such variables can be chosen either from the 
straight $\{ g_{\alpha\beta}, \pi^{\mu\nu} \}$, or dual $\{ g^{\alpha\beta}, \pi_{\mu\nu} \}$  sets of 
symplectic dynamical variables which always arise (and complete each other) in any Hamiltonian formulation 
developed for the coupled system of tensor fields. By applying canonical transformation(s) of dynamical 
variables we reduce the canonical Hamiltonian $H_C$ to its natural form. The natural form of canonical 
Hamiltonian provides numerous advantages in actual applications to the metric GR, since the general theory 
of dynamical systems with such Hamiltonians is well developed. Furthermore, many analytical and numerically 
exact solutions have been found and described in detail for dynamical systems with the Hamiltonians already 
reduced to their natural forms. In particular, reduction of the canonical Hamiltonian $H_C$ to its natural 
form allows one to derive the Jacobi equation for the free gravitational field(s), which takes a particularly 
simple form. \\

\noindent 
PACS number(s): 04.20.Fy and 11.10.Ef \\

\end{abstract}

\maketitle

%

\section{Introduction}

In 1958 Dirac published his famous Hamiltonian formulation  of the metric General Relativity (also 
the metric gravity, or GR, for short) \cite{Dir58}. Since then and for a very long time that Dirac's 
formulation was known as the only correct Hamiltonian approach ever developed for the metric gravity. 
In particular, only by using this Hamiltonian formulation and all primary and secondary constraints 
arising in this Dirac's approach, one is able to restore the complete and correct gauge symmetry 
(diffeomorphism) of the free gravitational field(s). A different Hamiltonian formulation of the metric 
gravity published earlier in \cite{PirSS} was overloaded with numerous mistakes, which can easily be 
found, e.g., in all secondary constraints derived in \cite{PirSS}. Moreover, some important steps of 
the complete Hamiltonian procedure, developed earlier by Dirac in \cite{Dir50}, were missing in 
\cite{PirSS}. For instance, the closure of Dirac procedure \cite{Dir50} was not demonstrated et al. 
In reality, it is impossible to show such a closure with wrong secondary constraints, but after 
reading \cite{PirSS} one can get an impression that authors did not understand why they need to do 
this, in principle. The complete and correct version of the Hamiltonian formulation of the metric 
gravity, originally proposed in \cite{PirSS}, was re-developed and substantially corrected only in 
2008 \cite{K&K} by Kiriushcheva and Kuzmin. Below, to respect this fact we shall call the Hamiltonian 
formulation of the metric GR developed in \cite{K&K} by the K$\&$K approach. This approach also 
allows one to restore the complete diffeomorphism as a correct and unique gauge symmetry of the free 
gravitational field. 

Note that after publication \cite{K&K} there were two different and non-contradictory Hamiltonian 
formulations of the metric gravity. Therefore, it was very interesting to investigate relations 
between these two approaches. In \cite{FK&K} we have shown that the original (or Dirac) formulation 
of the metric GR and alternative K$\&$K-formulation are related to each other by a canonical 
transformation of dynamical variables of the problem, i.e., by a transformation of the generalized 
`coordinates' $g_{\alpha\beta}$ and corresponding `momenta' $\pi^{\mu\nu}$. Furthermore, such a 
canonical transformation in metric GR always has some special and relatively simple form (more 
details can be found in \cite{FK&K}). After an obvious success of our analysis in \cite{FK&K} the 
following question has suddenly arose: is it possible to derive another canonical transformation of 
dynamical variables in the metric gravity which can reduce the canonical Hamiltonian $H_C$ of the 
metric GR derived in \cite{K&K} to some relatively simple forms, e.g., to the forms which are well 
known in classical mechanics? If the answer is `Yes', then we can use the solutions known for 
classical Hamiltonian systems to solve new gravitational problems, rigorously predict properties of 
certain gravitational systems, etc. Below, to answer this question we present the new canonical 
transformation of dynamical variables, i.e., generalized coordinates and momenta, in the metric 
General Relativity. This new canonical transformation is also a very special and unique, since it 
reduces the canonical Hamiltonian $H_C$ of metric GR to the natural form which is almost identical 
to the natural form of many `regular' Hamiltonians already known in analytical mechanics of the 
potential dynamical systems. 

In this paper we want to formulate and discuss all essential principles of the Hamiltonian 
formulation(s) of the metric General Relativity. To achieve this goal, in the next two Sections we 
introduce the $\Gamma - \Gamma$ Lagrangian ${\cal L} \equiv {\cal L}_{\Gamma - \Gamma}$ of the 
metric General Relativity. By using this Lagrangian ${\cal L}$ we define the corresponding momenta 
$\pi^{\alpha\beta}$. At the next stage we apply the Legendre transformation to exclude velocities, 
obtain the primary constraints and construct the canonical $H_C$ and total $H_t$ Hamiltonians of 
the metric General Relativity. All formulas and expressions derived in next two Sections and even 
logic used there are pretty standard for any Hamiltonian formulation of the metric GR. Moreover, 
some of these formulas were derived and discussed in a number of earlier studies (see, e.g., 
\cite{K&K} and \cite{Fro1}). Nevertheless, the two following Sections are important to make and 
keep this study completely independent of other publications and united by a central idea to 
illustrate the power of canonical transformations for Hamiltonian systems. 

The fundamental Poisson brackets of the metric GR are defined and calculated in Section III. These 
brackets are the main working tools to obtain accurate (and even exact) solutions of many 
gravitational problems and perform research of various Hamiltonian gravitational systems, including 
gravitational field(s) defined in the metric General Relativity. In particular, our fundamental and 
secondary Poisson brackets are used to investigate a few fundamental problems currently known in 
metric GR (see, Section IV). Section V is the central part of this study, since here the canonical 
Hamiltonian $H_C$ of the metric GR is reduced to its natural form. Here we also illustrate a number 
of advantages of the normal form of the canonical Hamiltonian $H_C$ for numerous problems known in 
the metric GR. A few directions for future development of the Hamiltonian formulations of metric GR 
are also discussed there. In Section VI we derive the Jacobi equation for the metric gravity by using 
our new dynamical and canonical variables $g_{\alpha\beta}$ and $P^{\mu\nu}$. Concluding remarks can 
be found in the last Section. There are also three Appendixes. In Appendix A we discuss relations 
between dynamical variables which are used in our and Dirac formulations of the metric General 
Relativity, while in the Appendix B we show that dynamical variables of modern geometro-dynamics are 
not (and cannot be) canonical variables of the metric GR. Appendix C contains explanation of some 
important `technical' details of our procedure which could not be included in the main text. 

Now, we need to introduce a few principal notations which are extensively used below. In particular, 
in this study the notation $g_{\alpha\beta}$ stands for the covariant components of the metric tensor 
(see, e.g., \cite{Kochin}) which are dimensionless values. The determinant of this metric tensor is 
$g$ which is the negative value, but $- g$ is always positive. It is assumed below that an arbitrary 
Greek index varies between 0 and $d - 1$, while an arbitrary Latin index varies between 1 and $d - 1$, 
where $d$ designates the total dimension of our space-time manifold ($d \ge 3$ \cite{X}). The 
quantities and tensors such as $B^{((\alpha \beta) \gamma | \mu \nu \lambda)}, I_{mnpq}$, etc, applied 
below, have been defined in earlier papers \cite{Dir58}, \cite{K&K}, \cite{FK&K} and \cite{Fro1}. In 
this study the definitions of all these quantities and tensors are exactly the same as in the mentioned 
papers and there is no need to repeat them. The short notations $g_{\alpha\beta,k}$ and 
$g_{\gamma\rho,0}$ are used below for the spatial and temporal derivatives, respectively, of the 
corresponding components of the metric tensor. Any expression which contains a pair of identical (or 
repeated) indexes, where one index is covariant and another index is contravariant, means summation over 
this `dummy' index. This convention is very useful and drastically simplifies many formulas derived in 
metric GR. 

\section{$\Gamma - \Gamma$ Lagrangian of the metric General Relativity}

In this Section we introduce the Lagrangian of the metric General Relativity. Formally, such a Lagrangian 
(or Lagrangian density) should coincide with the integrand in the Einstein-Hilbert integral-action 
$L_{EH}$ (see, e.g., \cite{LLTF} and \cite{Carm}) which equals to the product of scalar curvature of the 
$d-$dimensional space $R = g^{\alpha\beta} R_{\alpha\beta}$ and $\sqrt{- g}$. Here $R_{\alpha\beta}$ is 
the Ricci tensor (in old books and papers (see, e.g., \cite{Kochin}), it was called the Einstein tensor)
\begin{eqnarray}
 R_{\alpha\beta} = \frac{\partial \Gamma^{\gamma}_{\alpha\beta}}{\partial x^{\gamma}} - \frac{\partial 
 \Gamma^{\gamma}_{\alpha\gamma}}{\partial x^{\beta}} + \Gamma^{\gamma}_{\alpha\beta} 
 \Gamma^{\lambda}_{\gamma\lambda} - \Gamma^{\lambda}_{\alpha\gamma} \Gamma^{\gamma}_{\beta\lambda} \; , 
 \; \label{LGG01} 
\end{eqnarray}
where $\Gamma^{\gamma}_{\alpha\beta} = \frac12 g^{\gamma\nu} \Bigl( \frac{\partial g_{\nu\alpha}}{\partial 
x^{\beta}} + \frac{\partial g_{\nu\beta}}{\partial x^{\alpha}} - \frac{\partial g_{\alpha\beta}}{\partial 
x^{\nu}} \Bigr)$ are the Cristoffel symbols (see, e.g., \cite{Carm} and \cite{Kochin}). It is easy to see 
that this Lagrangian, which is often called the Einstein-Hilbert Lagrangian, contains a few derivatives of 
the second order $\frac{\partial^{2} g_{\alpha\beta}}{\partial x^{\gamma} \partial x^{\lambda}}$ and cannot 
be used directly in the principle of least action. However, all these derivatives of the second order are 
included in the Lagrangian $L_{EH}$ only as a linear combination with the constant coefficients, which equal 
+1, or -1. Because of such a linearity the invariant integral $\int R \sqrt{- g} d\Omega$ (gravitational 
action) can be transformed by means of Gauss theorem to the integral which does not include any second order 
derivatives. Indeed, we can represent this gravitational action integral in the form 
\begin{eqnarray}
 \int R \sqrt{- g} d\Omega = \int g^{\alpha\beta} \Bigl( \Gamma^{\lambda}_{\alpha\gamma} 
 \Gamma^{\gamma}_{\beta\lambda} - \Gamma^{\gamma}_{\alpha\beta} \Gamma^{\lambda}_{\gamma\lambda}\Bigr) 
 \sqrt{- g} d\Omega + \int \frac{\partial \Bigl[\sqrt{- g} \Bigl( g^{\alpha\beta} 
 \Gamma^{\gamma}_{\alpha\beta} - g^{\alpha\gamma} 
 \Gamma^{\beta}_{\alpha\beta} \Bigr)\Bigr]}{\partial x^{\gamma}} d\Omega \; , \label{LGG10} 
\end{eqnarray}
where the integrand of the first integral on the right-hand side of this equation contains only 
components of the metric tensor and their first-order derivatives, while the second integral has 
the form of a divergence of the vector-like quantity $\sqrt{- g} \Bigl( g^{\alpha\beta} 
\Gamma^{\gamma}_{\alpha\beta} - g^{\alpha\gamma} \Gamma^{\beta}_{\alpha\beta} \Bigr)$. This second 
integral is transformed by applying Gauss theorem into an integral over a hypersurface surrounding 
the $d-$dimensional volume over which the integration is carried out in other two integrals. When 
we vary the gravitational action, the variation of this (second) term on the the right-hand side of 
the last equation vanishes, since in respect to the principle of least action, the variation of the 
(varied) field at the limits of the region of integration must be zero. 

Now, we may write for the variations of all terms in the previous equation 
\begin{eqnarray}
 \delta \int R \sqrt{- g} d\Omega = \delta \int L_{\Gamma-\Gamma} d\Omega \; \; \; {\rm or} \; \; 
 \; \frac{\delta \Bigl(\int R \sqrt{- g} \Bigr)}{\delta g_{\mu\nu}} = \frac{\delta 
 L_{\Gamma-\Gamma}}{\delta g_{\mu\nu}} \; , \; \label{LGG}
\end{eqnarray}
where the notation $\delta$ means variation, while the notation $\frac{\delta F}{\delta g_{\mu\nu}}$ 
means the variational derivative (or Lagrange derivative). Also in this equation the symbol 
$L_{\Gamma-\Gamma} = \sqrt{- g} g^{\alpha\beta} \Bigl( \Gamma^{\lambda}_{\alpha\gamma} 
\Gamma^{\gamma}_{\beta\lambda} - \Gamma^{\gamma}_{\alpha\beta} \Gamma^{\lambda}_{\gamma\lambda} \Bigr)$ 
stands for the `regular' $\Gamma - \Gamma$ Lagrangian density (or Lagrangian, for short) of the metric 
gravity which plays a central role in any Hamiltonian approach developed the metric gravity. As follows 
from this equation the variational derivative of the $L_{\Gamma-\Gamma}$ Lagrangian is the true tensor, 
while the original $L_{\Gamma-\Gamma}$ Lagrangian is not a true scalar. Other properties of the 
$L_{\Gamma-\Gamma}$ Lagrangian are mentioned below. The equality, Eq. (\ref{LGG}), expresses the fact 
that we have replaced the `singular' Einstein-Hilbert Lagrangian ($L_{EH} = \sqrt{- g} R$) by the 
`regular' $\Gamma - \Gamma$ Lagrangian $L_{\Gamma-\Gamma} = \sqrt{- g} g^{\alpha\beta} \Bigl( 
\Gamma^{\lambda}_{\alpha\gamma} \Gamma^{\gamma}_{\beta\lambda} - \Gamma^{\gamma}_{\alpha\beta} 
\Gamma^{\lambda}_{\gamma\lambda} \Bigr)$ which is variationally equivalent to the original 
Einstein-Hilbert Lagrangian and contains no second order derivative. This $\Gamma - \Gamma$ Lagrangian 
is also written in the following form
\begin{eqnarray}
 {\cal L}_{\Gamma - \Gamma} = \frac14 \sqrt{-g} B^{\alpha\beta\gamma\mu\nu\rho} \Bigl(\frac{\partial 
 g_{\alpha\beta}}{\partial x^{\gamma}}\Bigr) \Bigl(\frac{\partial g_{\mu\nu}}{\partial x^{\rho}}\Bigr) 
 = \frac14 \sqrt{-g} B^{\alpha\beta\gamma\mu\nu\rho} g_{\alpha\beta,\gamma} g_{\mu\nu,\rho} \; , \; 
 \label{eq05} 
\end{eqnarray}
where 
\begin{eqnarray}
 B^{\alpha\beta\gamma\mu\nu\rho} &=& g^{\alpha\beta} g^{\gamma\rho} g^{\mu\nu} - g^{\alpha\mu} 
 g^{\beta\nu} g^{\gamma\rho} + 2 g^{\alpha\rho} g^{\beta\nu} g^{\gamma\mu} - 2 g^{\alpha\beta} 
 g^{\gamma\mu} g^{\nu\rho} \; \; \label{Bcoef}
\end{eqnarray}
is a homogeneous cubic function of the contravariant components of the metric tensor $g^{\alpha\beta}$. 
Below, we deal with the $\Gamma - \Gamma$ Lagrangian only. Therefore, to simplify the following formulas 
we shall designate this $L_{\Gamma-\Gamma}$ Lagrangian $L_{\Gamma-\Gamma}$ by the letter $L$, i.e., 
everywhere below $L = L_{\Gamma-\Gamma}$.  

By using this $\Gamma - \Gamma$ Lagrangian we need to define the corresponding momenta. First, note that 
in this study the covariant components of the metric tensor $g_{\alpha\beta}$ are chosen as the straight 
set of coordinates for the Hamiltonian formulation(s) of the metric GR. The contravariant components of 
the metric tensor $g^{\alpha\beta}$ form a different set of dual coordinates. Note also that in the 
right-hand side of this formula, Eq.(\ref{eq05}), the short notation $g_{\alpha\beta,\gamma}$ designates 
the partial derivatives $\frac{\partial g_{\alpha\beta}}{\partial x^{\gamma}}$ in respect to the spatial 
($g_{\alpha\beta,k}$) and temporal ($g_{\alpha\beta,0}$) coordinates. The partial temporal derivatives 
$g_{0 \sigma,0} (= g_{\sigma 0,0})$ of the $g_{0 \sigma}$ components are often called the 
$\sigma$-velocities. In reality, to derive the closed formula for the Hamiltonian of metric GR we need a 
slightly different form of the $\Gamma - \Gamma$ Lagrangian where all temporal derivatives (or 
time-derivatives) are explicitly separated from other derivatives (see, e.g., \cite{K&K})  
\begin{eqnarray}
  {\cal L} = \frac14 \sqrt{-g} B^{\alpha\beta 0\mu\nu 0} g_{\alpha\beta,0} g_{\mu\nu,0} + \frac12 
  \sqrt{-g} B^{(\alpha\beta 0|\mu\nu k)} g_{\alpha\beta,0} g_{\mu\nu,k} + \frac14 \sqrt{-g} 
  B^{\alpha\beta k \mu\nu l} g_{\alpha\beta,k} g_{\mu\nu,l} \; , \; \label{eq51}
\end{eqnarray}
where the notation $B^{(\alpha\beta\gamma|\mu\nu\rho)}$ means a `symmetrical' 
$B^{\alpha\beta\gamma\mu\nu\rho}$ quantity which is symmetrized in respect to the permutation of 
two groups of indexes, i.e.,
\begin{eqnarray}
 B^{(\alpha\beta\gamma|\mu\nu\rho)} &=& \frac12 \Bigl( B^{\alpha\beta\gamma\mu\nu\rho} + 
 B^{\mu\nu\rho\alpha\beta\gamma} \Bigr) = g^{\alpha\beta} g^{\gamma\rho} g^{\mu\nu} - g^{\alpha\mu} 
 g^{\beta\nu} g^{\gamma\rho} \nonumber \\ 
 &+& 2 g^{\alpha\rho} g^{\beta\nu} g^{\gamma\mu} - g^{\alpha\beta} g^{\nu\rho} g^{\gamma\mu} - 
 g^{\alpha\rho} g^{\beta\gamma} g^{\mu\nu}  \; . \; \label{eq52}
\end{eqnarray}

By using the Lagrangian ${\cal L}$, Eq.(\ref{eq51}), and standard definition of momentum as a partial 
derivative of the Lagrangian in respect to the corresponding velocity (see, e.g., \cite{Dir64}), one 
obtains the explicit formulas for all components of the tensor of momentum $\pi^{\gamma\sigma}$   
\begin{eqnarray}
  \pi^{\gamma\sigma} = \frac{\partial {\cal L}}{\partial g_{\gamma\sigma,0}} = \frac{1}{2} \sqrt{-g} 
  B^{((\gamma\sigma) 0|\mu\nu 0)} g_{\mu\nu, 0} + \frac{1}{2} \sqrt{-g} B^{((\gamma\sigma) 0|\mu\nu k)} 
  g_{\mu\nu, k} \; \; . \; \label{mom}
\end{eqnarray}
The first term in the right-hand side of this equation can be written in the form 
\begin{eqnarray}
 \frac{1}{2} \sqrt{-g} B^{((\gamma\sigma)0|\mu\nu 0)} g_{\mu\nu, 0} = \frac{1}{2} \sqrt{-g} g^{00} 
 E^{\mu\nu\gamma\sigma} g_{\mu\nu, 0} \; , \; \; \label{B}
\end{eqnarray}
where the notations $E^{\mu\nu\gamma\sigma}$ and $e^{\mu \nu}$ stands for the Dirac tensors, which are
\begin{eqnarray}
 E^{\mu \nu \gamma \rho} = e^{\mu \nu} e^{\gamma \rho} - e^{\mu \gamma} e^{\nu \rho} \; \; , \; \; 
 {\rm and} \; \; \; e^{\mu \nu} = g^{\mu \nu} - \frac{g^{0 \mu} g^{0 \nu}}{g^{00}} \; \; \; \label{E}  
\end{eqnarray}
and it is easy to check that $E^{\mu\nu\gamma\sigma} = E^{\gamma\sigma\mu\nu}$ and $e^{\mu \nu} = 
e^{\nu \mu}$. Also, as follows directly from the formula, Eq. (\ref{E}), the tensor $e^{\mu \nu}$ equals 
zero, if either index $\mu$, or index $\nu$ (or both) equals zero. The same statement is true for the 
Dirac $E^{\mu\nu\gamma\sigma}$ tensor, i.e., $E^{0\nu\gamma\sigma} = 0, E^{\mu 0\gamma\sigma} = 0, 
E^{\mu\nu 0\sigma} = 0$ and $E^{\mu\nu\gamma 0} = 0$. The $E^{pqkl}$ quantity is called the space-like 
Dirac tensor of the fourth rank. Note that all components of this space-like tensor $E^{p q k l}$ are 
not equal zero. Furthermore, the space-like tensor $E^{p q k l}$ is a positively-defined and invertable 
tensor. Its inverse space-like tensor $I_{m n p q}$ is also positively-defined and invertable space-like 
tensor of the fourth rank which is written in the form \cite{K&K} 
\begin{equation}
 I_{m n q p} = \frac{1}{d - 2} g_{m n} g_{p q} -  g_{m p} g_{n q} \; \; . \; \label{I}
\end{equation}
This tensor plays a very important role in our Hamiltonian analysis (see below). Now we can write 
$I_{m n p q} E^{p q k l} = g^{k}_{m} g^{l}_{n} = \delta^{k}_{m} \delta^{l}_{n}$, where the 
$g^{\alpha}_{\beta} = \delta^{\alpha}_{\beta}$ tensor is the substitution tensor \cite{Kochin}, while 
the symbol $\delta^{\alpha}_{\beta}$ denotes the Kroneker delta (it equals zero for all possible 
indexes, unless $\alpha = \beta$, when its numerical value equals unity). 

First, let us consider the `regular' case when in Eq.(\ref{mom}) $\gamma = p$ and $\sigma = q$. In 
this case one finds the following formula for space-like components of the momentum tensor
\begin{eqnarray}
  \pi^{pq} = \frac{\partial {\cal L}}{\partial g_{p q,0}} = \frac{1}{2} \sqrt{-g} 
  B^{((p q) 0|\mu\nu 0)} g_{\mu\nu,0} + \frac{1}{2} \sqrt{-g} B^{((p q) 0|\mu\nu k)} 
  g_{\mu\nu, k} \; \; \label{momenta}
\end{eqnarray}
for each pair of $(pq)-$indexes (or $(mn)-$indexes). The tensor in the right-hand side of this 
equation is invertable and the velocity $g_{m n, 0}$ is explicitly expressed as the linear 
function (or linear combination) of the space-like components $\pi^{pq}$ of momentum tensor:
\begin{eqnarray}
 g_{mn, 0} &=& \frac{1}{g^{00}} \Bigl( \frac{2}{\sqrt{-g}} I_{m n p q} \pi^{pq} - I_{m n p q} 
 B^{((pq) 0|\mu\nu k)} g_{\mu\nu, k} \Bigr) = \frac{1}{g^{00}} I_{m n p q} \Bigl( 
 \frac{2}{\sqrt{-g}} \pi^{pq} \nonumber \\
 &-& B^{((pq) 0|\mu\nu k)} g_{\mu\nu, k} \Bigr) \; \; , \; \label{veloc}
\end{eqnarray}
where the Dirac tensor $I_{m n p q}$ is defined by Eq.(\ref{I}). As follows from Eqs.(\ref{momenta}) 
and (\ref{veloc}) for the space-like components of metric tensor $g_{pq}$ and corresponding momenta 
$\pi^{mn}$ one finds no principal difference with the classical dynamical systems, which have 
Lagrangians written as the quadratic functions of the velocities. Indeed, for such systems the 
corresponding space-like components of momenta and corresponding velocities are related to each other 
by a few simple, linear equations. In the metric General Relativity, however, even for spatial 
components of momenta $\pi^{pq}$ and velocities $g_{pq,0}$ such relations take the multi-dimensional, 
matrix form. This means that one space-like component of momenta $\pi^{mn}$ depends upon quasi-linear 
combination \cite{QL} of different velocities $g_{pq,0}$ (and vice versa). Nevertheless, such a matrix 
definition of momenta works very well and allows one to develop a complete and non-contradictive 
Hamiltonian approach for the metric GR.

In the second `non-regular' (or singular) case, when $\gamma = 0$ in Eq.(\ref{mom}), the first term 
in the right-hand side of Eq.(\ref{mom}) equals zero and this equation takes the from 
\begin{eqnarray}
  \pi^{0\sigma} = \frac{\partial {\cal L}}{\partial g_{0\sigma,0}} = \frac{1}{2} \sqrt{-g} 
  B^{((0\sigma) 0|\mu\nu k)} g_{\mu\nu, k} \; \; , \; \label{constr}
\end{eqnarray}
which contains no velocity et al. This equation, Eq.(\ref{constr}), determines the momentum 
$\pi^{0\sigma}$ as a polynomial (cubic) functions of the contravariant components of the metric 
tensor $g^{\alpha\beta}$ and a linear function of the both $\sqrt{- g}$ value and spatial 
derivatives $g_{\mu\nu, k}$ of the covariant components of metric tensor $g_{\mu\nu}$. It is 
clear that such a situation cannot be found neither in classical mechanics, nor in quantum 
mechanics of arbitrary systems of particles. However, for actual physical fields similar 
situations arise quite often. The physical meaning of Eq.(\ref{constr}) is simple and can be 
expressed in the following words. The functions
\begin{eqnarray}
   \phi^{0\sigma} = \pi^{0\sigma} - \frac{1}{2} \sqrt{-g} B^{((0\sigma) 0|\mu\nu k)} 
   g_{\mu\nu, k} \; \; , \;  \label{primary}
\end{eqnarray}
where $\sigma = 0, 1, \ldots, d - 1$, must be equal zero during any actual physical motions 
(or time-evolution) of the gravitational field. In \cite{Dir50} Dirac proposed to write such 
equalities in the symbolic form $\phi^{0\sigma} \approx 0$ and called these $d$ functions 
$\phi^{0\sigma}$ (for $\sigma = 0, 1, \ldots, d - 1$), which are defined by Eq.(\ref{primary}), 
the primary constraints (see, e.g., \cite{Dir58}, \cite{Dir50}, \cite{Dir64}, \cite{Berg1} and 
\cite{Berg2}).

\section{Canonical and total Hamiltonians of metric General Relativity}

Now, by applying the Legendre transformation to the known $\Gamma - \Gamma$ Lagrangian 
${\cal L}$, of the metric GR, Eq. (\ref{eq51}), and excluding all space-like field-velocities 
$g_{mn,0}$ we can derive the explicit formulas for the total and canonical Hamiltonians of the 
metric GR \cite{Hamlt}. In particular, the total Hamiltonian $H_t$ of the gravitational field 
in metric GR derived from the $\Gamma - \Gamma$ Lagrangian ${\cal L}$, Eq. (\ref{eq05}), is 
written in the form 
\begin{eqnarray}
  H_t = \pi^{\alpha\beta} g_{\alpha\beta,0} - {\cal L} = H_C + g_{0\sigma,0} \phi^{0\sigma} 
  \; \; , \; \label{eq1}
\end{eqnarray}
where $\phi^{0\sigma} = \pi^{0\sigma} - \frac{1}{2}\sqrt{-g} B^{\left( \left(0\sigma\right) 
0\mid\mu\nu k\right)} g_{\mu\nu,k}$ are the primary constraints, while $g_{0\sigma,0}$ are the 
corresponding $\sigma-$velocities and $H_C$ is the canonical Hamiltonian of metric GR
\begin{eqnarray}
 & &H_C = \frac{1}{\sqrt{-g} g^{00}} I_{mnpq} \pi^{mn} \pi^{pq} - \frac{1}{g^{00}} I_{mnpq} 
 \pi^{mn} B^{(p q 0|\mu \nu k)} g_{\mu\nu,k} \label{eq5} \\
 &+& \frac14 \sqrt{-g} \Bigl[ \frac{1}{g^{00}} I_{mnpq} B^{((mn)0|\mu\nu k)} 
 B^{(pq0|\alpha\beta l)} - B^{\mu\nu k \alpha\beta l}\Bigr] g_{\mu\nu,k} g_{\alpha\beta,l} 
 \; \; , \; \nonumber
\end{eqnarray}
which does not contain any primary constraint $\phi^{0\sigma}$. In contrast with $H_C$ the total 
Hamiltonian $H_t$, Eq. (\ref{eq1}) includes all $d$ primary constraints $\phi^{0\sigma}$, where 
$\sigma = 0, 1, \ldots, d - 1$. It should be emphasized again that these primary constraints 
arise during our transition from the $\Gamma - \Gamma$ Lagrangian ${\cal L}$, Eq. (\ref{eq05}), 
to the Hamiltonians $H_t$ and $H_C$, since the $\Gamma - \Gamma$ Lagrangian ${\cal L}$ is a 
linear (not quadratic!) function of all $d$ values $g_{0\sigma,0}$, or $\sigma-$velocities. The 
total and canonical Hamiltonians $H_t$ and $H_C$ are the scalar functions defined in the $d (d 
+ 1)-$dimensional phase space $\Bigl\{ g_{\alpha\beta}, \pi^{\mu\nu} \Bigr\}$, where all 
components of the metric $g_{\alpha\beta}$ and momentum $\pi^{\mu\nu}$ tensors have been chosen 
as the basic dynamical variables. Such a  $d (d + 1)-$dimensional phase space is, in fact, a 
symplectic space and the corresponding symplectic structure is determined by the Poisson brackets 
between its basic dynamical variables, i.e., coordinates $g_{\alpha\beta}$ and momenta 
$\pi^{\mu\nu}$. 

To make the next step we need to define the Poisson brackets (or PB, for short) which are 
absolutely crucial for creation, development and applications of Hamiltonian approaches for 
arbitrary, in principle, physical systems of particles and fields, including the metric General 
Relativity. From now on we shall consider only Hamiltonian approaches (in metric GR) which are 
canonically related either to the K$\&$K-approach \cite{K&K}, or to the Dirac approach 
\cite{Dir58}. Note again that these two Hamiltonian formulations are canonically related to each 
other (for more details, see \cite{FK&K}). Therefore, it is possible to obtain and present the 
basic (or fundamental) set of Poisson brackets only for one of these two Hamiltonian formulations, 
e.g., for the K$\&$K-approach. Analogous Poisson brackets for other Hamiltonian formulations of 
metric GR can be derived from these basic (or fundamental) values known in the K$\&$K-approach. 
The basic Poisson brackets between $\frac{d(d + 1)}{2}$ components of the momentum tensor 
$\pi^{\mu\nu}$ and $\frac{d(d + 1)}{2}$ `coordinates' $g_{\alpha\beta}$ in the K$\&$K-approach 
are \cite{K&K}
\begin{eqnarray}
  [ g_{\alpha\beta}, \pi^{\mu\nu}] = - [ \pi^{\mu\nu}, g_{\alpha\beta}] = g_{\alpha\beta} 
  \pi^{\mu\nu} - \pi^{\mu\nu} g_{\alpha\beta} = \frac12 \Bigl(g^{\mu}_{\alpha} g^{\nu}_{\beta} 
  + g^{\nu}_{\alpha} g^{\mu}_{\beta}\Bigr) = \frac12 \Bigl(\delta^{\mu}_{\alpha} 
  \delta^{\nu}_{\beta} + \delta^{\nu}_{\alpha} \delta^{\mu}_{\beta}\Bigr) = 
  \Delta^{\mu\nu}_{\alpha\beta} \; \; , \; \label{eq15} 
\end{eqnarray}
where $g^{\mu}_{\alpha} = \delta^{\mu}_{\alpha}$ is the substitution tensor \cite{Kochin} and symbol 
$\delta^{\mu}_{\beta}$ is the Kronecker delta, while the notation $\Delta^{\mu\nu}_{\alpha\beta}$ 
stands for the gravitational (or tensor) delta-symbol. All other fundamental Poisson brackets between 
basic dynamical variables of the metric GR equal zero identically, i.e., $[ g_{\alpha\beta}, 
g_{\mu\nu}] = 0$ and $[ \pi^{\alpha\beta}, \pi^{\mu\nu}] = 0$. Thus, for the $d-$dimensional metric 
gravity one finds $N_P = \frac{d^{2}(d + 1)^{2}}{2}$ different Poisson brackets, but many of these 
brackets equal zero identically, if these dynamical variables are canonical \cite{canon}. In our case 
this means that all our momenta $\pi^{\mu\nu}$ have been defined correctly. The set of $N_P$ Poisson 
brackets has a fundamental value, since these PB define the unique symplectic structure directly 
related to the Riemanian structure of the original $d (d + 1)$-dimensional tensor phase space $\{ 
g_{\alpha\beta}, \pi^{\mu\nu} \}$ and to the original metric tensor $g_{\alpha\beta}$. General 
properties of Poisson brackets and their symmetries are discussed, e.g., in \cite{Gant} - \cite{GF}. 
For any Hamiltonian dynamical system all values and functions must be expressed in terms of the basic 
dynamical variables, i.e., in terms of generalized coordinates and momenta. Furthermore, all 
arithmetical, mathematical and other operations between such values and functions must be reduced to 
the Poisson brackets. Analytical computations of the Poisson brackets is the only actual tool and 
language of the Hamiltonian theory.   

\section{Poisson brackets}

The $N_P$ Poisson brackets mentioned above are sufficient to operate successfully in any correct 
Hamiltonian approach developed for the metric GR. However, in many applications it is crucially 
important to determine other Poisson brackets, which are often called the secondary PB. The 
secondary PB are calculated between different analytical functions of the basic dynamical 
variables, i.e., coordinates and momenta, and these PB always appear in actual calculations. In 
general, it is difficult and time-consuming to derive the explicit formulas for such secondary 
PB every time when you need them. Furthermore, in actual applications one usually needs to 
determine a few hundreds of different Poisson brackets. Here we present a number of secondary 
Poisson brackets which are sufficient for our purposes in this study. The first additional group 
of secondary Poisson brackets is   
\begin{eqnarray}
 [ g^{\alpha\beta}, \pi^{\mu\nu}] = - \frac12 \Bigl( g^{\alpha\mu} g^{\beta\nu} + g^{\alpha\nu} 
 g^{\beta\mu} \Bigr) = [ \pi^{\mu\nu}, g^{\alpha\beta}] \; \; {\rm and} \; \; [ g^{\alpha\beta}, 
 g_{\mu\nu}] = 0 \; , \label{eq151} 
\end{eqnarray}
which include the contravariant components of the metric tensor $g^{\alpha\beta}$. These Poisson 
brackets are of great interest, since our canonical and total Hamiltonians (see above) are 
overloaded with the contravariant components of the metric tensor. 
  
The second set of additional Poisson brackets arises, if one explicitly introduces the dual system 
of dynamical variables $\{ g^{\alpha\beta}, \pi_{\mu\nu}\}$ which always exists for any tensor 
Hamiltonian system. In general, to create the truly correct, covariant and non-contradictory 
Hamiltonian formulation for some dynamical system of tensor fields it is much better to deal 
(instantly) with the two different $d (d + 1)-$dimensional sets of dynamical variables: (a) the 
straight set $\{ g_{\alpha\beta}, \pi^{\mu\nu}\}$, and (b) the corresponding dual set $\{ 
g^{\alpha\beta}, \pi_{\mu\nu}\}$. Applications of the two sets of dynamical variables makes our 
Hamiltonian formulation complete and physically transparent. The Poisson brackets between all 
dynamical variables from these two sets must be derived and carefully checked for non-contradictory. 
The Poisson brackets for the dual set of dynamical variables $\{ g^{\alpha\beta}, \pi_{\mu\nu}\}$
\begin{eqnarray}
  [ g_{\alpha\beta}, \pi_{\mu\nu}] = \frac12 \Bigl( g_{\alpha\mu} g_{\beta\nu} + g_{\alpha\nu} 
  g_{\beta\mu} \Bigr) \; \; {\rm and} \; \;
  [ g^{\alpha\beta}, \pi_{\mu\nu}] = - \frac12 \Bigl( g^{\alpha}_{\mu} g^{\beta}_{\nu} + 
  g^{\alpha}_{\nu} g^{\beta}_{\mu} \Bigr) = - \Delta^{\alpha\beta}_{\mu\nu} \; \; \label{eq153} 
\end{eqnarray}
and also $[ g^{\alpha\beta}, g^{\mu\nu}] = 0$ and $[ g_{\alpha\beta}, g^{\mu\nu}] = 0$. Another 
Poisson bracket which we want to present here is 
\begin{equation}
  [ \pi_{\alpha\beta}, \pi^{\mu\nu}] = \frac12 \Bigl( \delta_{\alpha}^{\mu} \pi_{\beta}^{\nu} + 
  \delta_{\alpha}^{\nu} \pi_{\beta}^{\mu} + \delta_{\beta}^{\mu} \pi_{\alpha}^{\nu} + 
  \delta_{\beta}^{\nu} \pi_{\alpha}^{\mu} \Bigr) \; \; \; , \; \; \label{pipi}
\end{equation}
where $\pi_{\kappa}^{\rho} = g^{\rho \lambda} \pi_{\lambda \kappa} = g_{\kappa \lambda} 
\pi^{\lambda \rho}$. This equality means that the co- and contra-covariant components of the 
momentum tensor do not commute with each other. If they commuted, then the direct and dual 
sets of dynamical variables in metric gravity would be equivalent and there would be no real 
need to apply the two sets of dynamical variables (straight and dual), since at each step of 
our procedure we can always express one set of dynamic variables in terms of another set and 
vice versa. However, this is not true for the metric GR. 

Let us present the following formula for the fundamental Poisson brackets which unites the both 
straight and dual sets of dynamical variables  
\begin{eqnarray}
  [ g_{\alpha\beta}, \pi^{\mu\nu}] = \Delta^{\mu\nu}_{\alpha\beta} = [ \pi_{\alpha\beta}, 
  g^{\mu\nu}]  \; \; . \; \label{eq1551} 
\end{eqnarray}
This beautiful formula includes two fundamental Poisson bracket(s) and clearly shows the 
differences which arise during transition from the straight set of canonical variables to 
analogous dual set. As follows from the formula, Eq.(\ref{eq155}), the truly dual system 
of dynamical variables (for the original $\{ g_{\alpha\beta}, \pi^{\mu\nu}\}$ system) must 
be $\{ -g^{\alpha\beta}, \pi_{\mu\nu}\}$ system rather then our dual $\{ g^{\alpha\beta}, 
\pi_{\mu\nu}\}$ system of variables introduced above.Below, we shall ignore this fact and 
consider the $\{ g_{\alpha\beta}, \pi^{\mu\nu}\} \rightarrow \{ g^{\alpha\beta}, 
\pi_{\mu\nu}\}$ transition as a canonical transformation of dynamical variables for our 
Hamiltonian formulation of the metric GR. Therefore, based on the general theory of 
canonical transformations in Hamiltonian systems described in \cite{Gant} we can write the 
following equality 
\begin{eqnarray} 
 \pi^{\mu\nu} \delta g_{\mu\nu} - H_t \delta t + \delta F = v \Bigl( \pi_{\mu\nu} 
 \delta g^{\mu\nu} - \overline{H}_t \delta t \Bigr) \; \; , \; \; \label{eq1553}
\end{eqnarray}
where $v$ is a real, non-zero number which is called the valence of this canonical transformation, 
while $F(t, g_{\alpha\beta}, \pi^{\gamma\sigma})$ is its generating function. The notations $H_t$ 
and $\overline{H}_t$ means the total Hamiltonians written in the both systems of dynamical 
variables, i.e., in the straight $\{ g_{\alpha\beta}, \pi^{\mu\nu}\}$ and dual $\{ g^{\alpha\beta}, 
\pi_{\mu\nu}\}$ systems of variables, respectively. It is clear that for such a canonical 
transformation we can use the same time $t$ (for both systems) and this transformation is univalent 
which means that $| v | = 1$. In our case we have found that in our case $v = - 1$. Furthermore, it 
is possible to show that for the $\{ g_{\alpha\beta}, \pi^{\mu\nu}\} \rightarrow \{ g^{\alpha\beta}, 
\pi_{\mu\nu}\}$ canonical transformation the generating function $F$ can be chosen in a very special 
form $F = S(t, g_{\mu\nu}, g^{\alpha\beta})$ which corresponds to the free canonical transformation(s). 
In this case the previous equation takes the form  
\begin{eqnarray} 
 \pi^{\mu\nu} \delta g_{\mu\nu} - H_t \delta t + \delta S(t, g_{\mu\nu}, g^{\alpha\beta}) 
 = v \Bigl( \pi_{\mu\nu} \delta g^{\mu\nu} - \overline{H}_t \delta t \Bigr) \; \; \; 
 \label{eq1555}
\end{eqnarray}
and three following equations are also obeyed (for $v = - 1$)
\begin{eqnarray} 
 \pi^{\mu\nu} = -\frac{\partial S}{\partial g_{\mu\nu}} \; \; , \; \; \pi_{\mu\nu} = 
 -\frac{\partial S}{\partial g^{\mu\nu}} \; \; {\rm and} \; \; \overline{H}_t + H_t = 
 \frac{\partial S}{\partial t} \; \; . \; \; \label{eq1557}
\end{eqnarray}
As follows from Eq. (\ref{eq1555}) the differential form $dS(t, g_{\mu\nu}, g^{\alpha\beta})$ is 
the total differential of the potential function $S(t, g_{\mu\nu}, g^{\alpha\beta})$ which is, in 
fact, the generating function of the free canonical transformations. This generating function 
always exist and can explicitly be constructed for an arbitrary Hamiltonian system of tensor 
fields. Also, these our equations open a short way to the Jacobi equation for the gravitational 
field in metric GR, but below we shall apply a different approach to solve this interesting 
problem (see Section VI below). 

The necessity to deal with the two sets of dynamical variables instantaneously is an important 
difference between Hamiltonian procedures developed for the affine vector spaces and Riemanian 
tensor spaces. It can be shown that only by dealing with the both straight and dual sets of 
dynamical variables we can guarantee the internal covariance and self-sustainability of our 
Hamiltonian approach developed for the metric GR. The fact that we need to operate with the 
both straight and dual systems of dynamical variables in any Hamiltonian formulation developed 
for tensor dynamical systems can be illustrated by the following reasoning. To construct the 
Hamiltonian formulation, we are free to choose either direct, or dual sets of dynamic variables. 
For any meaningful physical theory, these Hamiltonian formulations must be equivalent, i.e., 
they must be connected to each other by a canonical transformation. In other words, the two sets 
of dynamical variables $\{g^{\alpha\beta}, \pi^{\rho\sigma} \}$ and $\{g^{\alpha\beta}, 
\pi_{\rho\sigma} \}$ are absolutely equivalent in order to develop the new Hamiltonian 
formulation(s) of the metric GR. The two newly arising Hamiltonian formulations are related to 
each other by a canonical transformation of variables (see above). 

A few following Poisson brackets which are also useful in actual calculations. Let $g (> 0)$ will 
be the determinant of the metric tensor $g_{\alpha\beta}$ and $F(g)$ is an arbitrary analytical 
function of $g$. In this notation one finds
\begin{eqnarray}
  [ F(g), \pi^{\alpha\beta}] = \Bigl( \frac{\partial F}{\partial g} \Bigr) g g^{\alpha\beta} 
  \; \; \; {\rm and} \; \; \; [ \sqrt{- g}, \pi^{\alpha\beta}] = - \frac{1}{2 \sqrt{- g}} g 
  g^{\alpha\beta} = \frac12 \sqrt{- g} g^{\alpha\beta} \; \; \label{eq154} 
\end{eqnarray}
for $F(g) = \sqrt{- g}$, if the determinant $g$ is negative. Analogously, for the $\pi_{\alpha\beta}$ 
momentum we obtain 
\begin{eqnarray}
  [ F(g), \pi_{\alpha\beta}] = \Bigl( \frac{\partial F}{\partial g} \Bigr) g g_{\alpha\beta} \; 
  \; \; {\rm and} \; \; \; [ \sqrt{- g}, \pi_{\alpha\beta}] = - \frac{1}{2 \sqrt{- g}} g 
  g_{\alpha\beta} = \frac12 \sqrt{- g} g_{\alpha\beta} \; \label{eq155} 
\end{eqnarray}
These formulas lead to the following expressions 
\begin{eqnarray}
  [ \frac{1}{\sqrt{- g}}, \pi^{\alpha\beta}] = - \frac{1}{2 \sqrt{- g}} g^{\alpha\beta} \; \; 
  \; {\rm and} \; \; \; [ \frac{1}{\sqrt{- g}}, \pi_{\alpha\beta}] = - \frac{1}{2 \sqrt{- g}} 
  g_{\alpha\beta} \; , \; \label{eq155a} 
\end{eqnarray}
which are important for our calculations performed in the next Sections. All other Poisson brackets, 
which are often needed in analytical calculations, can be determined with the use of our PB presented 
in Eqs.(\ref{eq15}) - (\ref{eq155a}). In general, analytical computations of a large number of Poisson 
brackets in any Hamiltonian formulation of metric GR is a very good exercise in tensor calculus. 

Another example is slightly more complicated and includes the tensor(s) $e^{\mu \nu}$ defined above. 
From the explicit formulas for the components of $e^{\mu \nu}$ tensor, Eq.(\ref{E}), one finds that 
only non-zero elements of this tensor are located in the space-like corner of the total $e^{\mu \nu}$ 
tensor. These non-zero elements form the space-like $e^{pq}$ tensor (or space-like part of the total 
$e^{\mu \nu}$ tensor) which is often called the space-like Dirac tensor of the second rank. For this 
tensor one easily finds the following useful relation
\begin{eqnarray}
  g_{\alpha\beta} e^{\alpha\beta} = g_{\alpha\beta} g^{\alpha\beta} - g_{\alpha\beta} 
  \Bigl(\frac{g^{\alpha 0} g^{\beta 0}}{g^{00}}\Bigr) = d - g_{\beta}^{0} \; 
  \frac{g^{\beta 0}}{g^{00}} = d - \frac{g^{00}}{g^{00}} = d - 1 = g_{mn} e^{mn} \; , 
  \; \label{d-1}
\end{eqnarray}
where $g_{\alpha\beta} g^{\alpha\beta} = d$ and $d$ is the total dimension of our space-time continuum. 
By using our formulas for the Poisson brackets obtained above we derive the two following expressions 
\begin{eqnarray}
 [ e^{pq}, \pi^{\alpha\beta}] &=& - \frac12 \Bigl( g^{p\alpha} g^{q\beta} + g^{p\beta} 
 g^{q\alpha} \Bigr) + \frac12 \Bigl( g^{0\alpha} g^{p\beta} + g^{0\beta} g^{p\alpha} \Bigr) 
 \Bigl(\frac{g^{0q}}{g^{00}}\Bigr) \nonumber \\
 &+& \frac12 \Bigl(\frac{g^{0p}}{g^{00}}\Bigr) \Bigl( g^{0\alpha} g^{q\beta} + g^{0\beta} 
 g^{q\alpha} \Bigr) - \frac{g^{0p} g^{q\alpha} g^{0\beta} g^{0q}}{(g^{00})^2} \; \label{e-tens}
\end{eqnarray}  
and 
\begin{eqnarray}
 [ e^{pq}, \pi_{\alpha\beta}] = - \Delta^{pq}_{\alpha\beta} + \Delta^{0 p}_{\alpha\beta} 
 \Bigl(\frac{g^{0q}}{g^{00}}\Bigr) + \frac12 \Bigl(\frac{g^{0p}}{g^{00}}\Bigr) 
 \Delta^{0 q}_{\alpha\beta} - \Delta^{0 0}_{\alpha\beta} \frac{g^{0p} g^{0q}}{(g^{00})^2} 
 \; . \label{e-tensA}
\end{eqnarray}  
Analytical formulas for these PB are important, since there were some ideas to use components of 
this space-like tensor $e^{pq}$ as the new$\frac{d(d - 1)}{2}$ canonical variables, or new 
coordinates, for another `advanced' Hamiltonian formulation of the metric GR. As follows from 
Eqs.(\ref{e-tens}) and (\ref{e-tensA}) the complexity of arising Poisson brackets makes this idea 
unworkable. 

\section{Applications to actual problems of metric gravity}

The knowledge of the fundamental and secondary Poisson brackets allows one to achieve a number of 
goals in the Hamiltonian formulation(s) of metric General Relativity. In particular, by using these 
Poisson brackets we can complete the actual Hamiltonian formulation of the metric GR. Another problem 
which can be solved with the use of our Poisson brackets is explicit derivation of the Hamilton 
equations of motion for actual gravitational field(s) which are often called the time-evolution 
equations. Also, by using these Poisson brackets one can find some new canonical transformations which 
are simplify either the canonical Hamiltonian $H_C$, or secondary constraints $\chi^{0\sigma}$ (they 
are defined below). Another important problem is the reduction of the canonical Hamiltonian $H_C$ to 
its natural form. The first two problems are briefly discussed in the next two subsections. These two 
problems were extensively investigated in earlier studies \cite{K&K}, \cite{FK&K} and \cite{Fro1}. 
Therefore, there is no need for us here to discuss formulation of these problems and repeat all 
formulas derived in those papers. Here we just want to illustrate how our formulas for Poisson brackets 
allow one to simplify analytical calculations of many difficult expressions. In contrast with this, the 
third problem (i.e., reduction of $H_C$ to its natural form) is one of the central parts of this study 
and we want to disclose all details of our computations. This problem with all details is described in 
the next Section. Another aim of this study is to derive Jacobi equation for the free gravitational 
field(s) in our new Hamiltonian formulation of the metric gravity. This problem is considered in 
Section VII.  

Let us complete the Hamiltonian formulation of the metric GR described above, by using the space-like 
momenta $\pi^{mn}$, its temporal components $\pi^{0\sigma}$ (or primary constraints $\phi^{0\sigma}$) 
and canonical Hamiltonian $H_C$ defined in Eq.(\ref{momenta}), Eq.(\ref{constr}) and Eq.(\ref{eq5}), 
respectively. First, we need to determine PB between the canonical Hamiltonian $H_C$, Eq.(\ref{eq5}), 
and primary constraints $\phi^{0\sigma}$, Eq.(\ref{primary}). This directly leads to the secondary 
constraints $\chi^{0\sigma} = [ H_C, \phi^{0\sigma} ]$, where $\sigma = 0, 1, \ldots, d - 1$ (see 
discussion in \cite{K&K}), since these secondary constraints $\chi^{0\sigma}$ do not equal zero 
identically. In Dirac procedure these $d$ secondary constraints $\chi^{0\sigma}$ become an integral 
part of the Hamilton formulation \cite{constr}. The explicit formulas for the secondary constraints 
$\chi^{0\sigma}$ are \cite{K&K} (see also \cite{Fro1}): 
\begin{eqnarray}
 \chi^{0\sigma} &=& -\frac{g^{0\sigma}}{2 \sqrt{-g} g^{00}} I_{mnpq} \pi^{mn} \pi^{pq} + 
 \frac{g^{0\sigma}}{2 g^{00}} I_{mnpq} \pi^{mn} U^{( pq0 \mid \mu\nu k )} g_{\mu\nu,k} + 
 \Bigl[ \pi_{,k}^{\sigma k} + \Bigl(\pi^{pk} e^{q \sigma} \label{eqn8} \\
 &-& \frac12 \pi^{pq} e^{k \sigma}) g_{pq,k} \Bigr] - \frac{\sqrt{-g}}{8} 
 \Bigl(\frac{g^{0\sigma}}{g^{00}} I_{mnpq} B^{((mn) 0 \mid \mu \nu k)} 
 B^{(pq0 \mid \alpha \beta t)} - g^{0\sigma} B^{\mu \nu k \alpha \beta t} \Bigr) 
 g_{\mu\nu,k} g_{\alpha\beta,t} \nonumber \\
 &+& \frac{\sqrt{-g}}{4 g^{00}} I_{mnpq} B^{((mn) 0 \mid\mu\nu k )} g_{\mu\nu,k} 
 g_{\alpha\beta,t} \Bigl[ g^{\sigma t} \Bigl( g^{00} g^{p \alpha} g^{q \beta} + g^{pq} 
 g^{0 \alpha} g^{0 \beta} - 2 g^{\alpha q} g^{0 p} g^{0 \beta} \Bigr) \nonumber \\
 &-& g^{\sigma p} \Bigl( 2 g^{00} g^{q \alpha} g^{t \beta} - g^{00} g^{\alpha \beta} g^{q t} 
 + g^{\alpha\beta} g^{0q} g^{0t} - 2 g^{q \alpha} g^{0 \beta} g^{0t} - 
 2 g^{t \alpha} g^{0 \beta} g^{0q} + 2 g^{qt} g^{0\alpha} g^{0\beta} \Bigr) \nonumber \\ 
 &+& g^{0\sigma} ( 2 g^{\beta t} g^{\alpha p} g^{0q} - 2 g^{p\alpha} g^{q\beta} g^{0t} 
 - 2 g^{pq} g^{t\beta} g^{0\alpha} + 2 g^{pt} g^{q\beta} 
 g^{0\alpha} + g^{pq} g^{\alpha\beta} g^{0t} - g^{tp} g^{\alpha\beta} g^{0q}) \Bigr] 
 \nonumber \\
 &-& \frac{\sqrt{-g}}{4} g_{\mu\nu,k} g_{\alpha\beta,t} \Bigl[ g^{\sigma t} 
 ( g^{\alpha\mu} g^{\beta\nu} g^{0k} + g^{\mu\nu} g^{\alpha t} g^{0\beta} - 2 
 g^{\mu\alpha} g^{k\nu} g^{0\beta} ) \nonumber \\
 &+& g^{0\sigma} ( 2 g^{\alpha t} g^{\beta\mu} g^{\nu k} - 3 g^{t\mu} g^{\nu k} 
 g^{\alpha\beta} - 2 g^{\mu\alpha} g^{\nu\beta} g^{kt} + g^{\mu\nu} g^{kt} 
 g^{\alpha\beta} + 2 g^{\mu t} g^{\nu\beta} g^{k\alpha}) \nonumber \\
 &+& g^{\sigma\mu} \Bigl( (g^{\alpha\beta} g^{\nu t} - 2 g^{\nu\alpha} g^{t\beta}) 
 g^{0k} + 2 ( g^{\beta\nu} g^{kt} - g^{\beta k} g^{t\nu}) g^{0\alpha} + ( 2 
 g^{k\beta} g^{\alpha t} - g^{\alpha\beta} g^{kt}) g^{0\nu}\Bigr) \Bigr] \nonumber \\
&-& \frac{\sqrt{-g} g^{00}}{2} E^{pqt\sigma} \Bigl( \frac{1}{g^{00}} I_{mnpq} 
B^{((mn)0 \mid \mu\nu k)} g_{\mu\nu,k} \Bigr)_{,t} - \frac{\sqrt{-g}}{2} B^{((\sigma 0) 
k \mid \alpha \beta t)} g_{\alpha\beta,kt} \; , \nonumber
\end{eqnarray}
where $U^{( pq 0 \mid \mu\nu k )}$ is the symmetrized form of the following expression 
\begin{eqnarray}
 U^{\alpha\beta 0 \mu\nu k} = B^{(\alpha\beta 0 \mid \mu \nu k)} - g^{0k} E^{\alpha \beta \mu \nu} 
 + 2 g^{0\mu} E^{\alpha \beta k \nu} \; \; \label{Sab}
\end{eqnarray}
and $\sigma = 0, 1, \ldots, d - 1$. The total number of primary and secondary constraints for 
this Hamilton formulation equals $d + d = 2 d$. Note also that all these primary and secondary 
constraints $\phi^{0\sigma}$ and $\chi^{0\sigma}$, where $\sigma = 0, 1, \ldots, d - 1$, are 
the first-class constraints \cite{Dir64}. In general, our formulas for Poisson brackets (or PB, 
for short) substantially simplify the whole process of derivation of the explicit formulas for 
the primary and secondary constraints and for PB between them. In particular, by using our 
Poisson brackets one can show that all Poisson brackets between primary constraints equal zero 
identically, i.e., $[ \phi^{0\lambda}, \phi^{0\sigma} ] = 0$, while $[ \phi^{0\lambda}, 
\chi^{0\sigma} ] = \frac12 g^{\lambda\sigma}$. The Poisson brackets between canonical Hamiltonian 
$H_C$ and secondary constraints $\chi^{0\sigma}$ are expressed as `quasi-linear' \cite{QL} 
combinations of the same secondary constrains $\chi^{0\sigma}$, i.e., we obtain
\begin{eqnarray}
 [ \chi^{0\sigma}, H_{c} ] &=& -\frac{2}{\sqrt{-g}} I_{mnpq} \pi^{mn} 
 \Bigl(\frac{g^{\sigma q}}{g^{00}}\Bigr) \chi^{0p} + \frac12 
 g^{\sigma k} g_{00,k} \chi^{00} + \delta_{0}^{\sigma} \chi_{,k}^{0k} \label{close} \\
 &+& \Bigl( -2 \frac{1}{\sqrt{-g}} I_{mnpk} \pi^{mn} \frac{g^{\sigma p}}{g^{00}} + 
 I_{mkpq} g_{\mu\nu,l} \frac{g^{\sigma m}}{g^{00}} 
 U^{(pq) 0 \mu\nu l} \Bigr)\chi^{0k} \nonumber \\
 &-& \Bigl( g^{0\sigma} g_{00,k} + 2 g^{n\sigma} g_{0n,k} + \frac{g^{n\sigma} 
 g^{0m}}{g^{00}} (g_{mn,k} + g_{km,n} - g_{kn,m}) \Bigr) \chi^{0k} \; , \nonumber
\end{eqnarray}
where $U^{(pq) 0 \mu\nu k}$ is the quantity $U^{\alpha\beta 0 \mu\nu k}$ from Eq.(\ref{Sab}) 
which is symmetrized upon all $p \leftrightarrow q$ permutations. The Poisson bracket, 
Eq.(\ref{close}), indicates that the Hamilton procedure developed for the metric GR in 
\cite{K&K} and \cite{FK&K} is closed (Dirac closure), i.e., the Poisson bracket $[ 
\chi^{0\sigma}, H_{c} ]$ does not lead to any tertiary, or other constraints of higher order(s). 
Analogously, the Poisson brackets between secondary constraints $[ \chi^{0\sigma}, 
\chi^{0\gamma}]$, where $\sigma \ne \gamma$ (if $\sigma = \gamma$, then this PB equals zero 
identically), are
\begin{eqnarray}
 [ \chi^{0\sigma}, \chi^{0\gamma} ] &=& [ \chi^{0\sigma}, [ \phi^{0\gamma}, H_{c} ]] = - [ 
 \phi^{0\gamma}, [ H_C, \chi^{0\sigma} ]] - [ H_C, [ \chi^{0\sigma}, \phi^{0\gamma} ]] 
 \nonumber \\ 
 &=& [ \phi^{0\gamma}, [ \chi^{0\sigma}, H_C ]] - \frac12 [ g^{\sigma\gamma}, H_C ] \; \; , 
 \; \label{chichi} 
\end{eqnarray} 
where the Poisson bracket $[ \chi^{0\sigma}, H_C ]$ is given by the formula, Eq. (\ref{close}). This 
formula also does not lead to any constraint of higher order (see discussion in \cite{Dir64}). This 
proves that the Hamiltonian system which includes the canonical Hamiltonian $H_C$ and all primary 
$\phi^{0\lambda}$ and secondary $\chi^{0\sigma}$ constraints \cite{constr} is closed (here $\lambda = 
0, 1, \ldots, d - 1$ and $\sigma = 0, 1, \ldots, d - 1$). The actual closure of the Dirac procedure 
\cite{Dir50} for the Hamiltonian formulation of the metric GR was shown for the first time in 
\cite{K&K}. Formally, the explicit demonstration of closure of the whole Dirac procedure \cite{Dir50} 
is the last and most important step for any Hamiltonian formulation of the metric GR \cite{Dir64}, 
\cite{Tyut}. However, in reality one needs to check one more condition which appears to be crucial for 
separation of the actual Hamiltonian formulations of the metric GR  from numerous quasi-Hamiltonian 
constructions developed in this area of gravitational research, since the end of 1950's. 

This additional condition follows from rigorous conservation of the gauge invariance (or symmetry) 
of the metric GR during transformations from the original $\Gamma - \Gamma$ Lagrangian to the 
Hamiltonian formulation. In other words, we cannot reduce (or increase) the gauge symmetry known 
for the free gravitational field(s) which is obeyed the original Einstein equations. Disappearance 
(or reduction) of the gauge invariance of the original problem simply means that our transformations 
to the Hamiltonian formulation are wrong and non-equivalent, or simply that they are not canonical. 
The formulas for the Hamiltonians $H_t, H_C$ presented above and explicit expressions for all primary 
and secondary constraints \cite{K&K}, \cite{Fro1} allow one to derive (with the use of Castellani 
procedure \cite{Cast}) the correct generators of gauge transformations, which directly and 
unambiguously lead to the diffeomorphism invariance \cite{K&K}. The diffeomorphism invariance is well 
known gauge symmetry (or gauge, for short) of the free gravitational field(s) which was discovered in 
early years of the metric GR (see, e.g., \cite{Carm} and references therein). In particular, this 
gauge symmetry directly leads (see, e.g., \cite{Tyut}) to the $d$ different Bianchi identities which 
are well known for the Ricci tensor since 1880 \cite{Voss}. These $d$ Bianchi identities for the Ricci 
tensor can be written in the following `tensor' form: 
\begin{eqnarray}
  \nabla_{\mu} \Bigl( R^{\alpha\beta} - \frac12 \; g^{\alpha\beta} \; R \Bigr) = 0 \; \;  , \; \; 
  {\rm where}  \; \; \; \nabla_{\mu} R^{\alpha\beta} = \frac{\partial R^{\alpha\beta}}{\partial 
  x^{\mu}} + R^{\alpha\lambda} \; \Gamma^{\beta}_{\lambda \mu}  + R^{\lambda\beta} 
  \Gamma^{\alpha}_{\lambda \mu} \; \; , \; \label{totderiv}
\end{eqnarray}
is the covariant tensor derivative and $\nu = 0, 1, \ldots, d - 1$. Below, we have shown that in ADM 
gravity it is impossible to derive the diffeomorphism as the actual gauge invariance of the metric GR. 
Therefore, one  cannot apply the Bianchi identities in the framework of ADM gravity. 

The Castellani procedure is based on the explicit derivation of generators of gauge transformations
which are unambiguously defined by the chain of first-class constraints \cite{K&K}. In general, we 
start from the primary first-class constraints and then construct the complete set of generators of 
gauge transformation(s). These primary constraints play the central role in the Castellani procedure, 
since each of these constraints generates a separate chain of gauge generators. Furthermore, during 
the actual motion of any constrained dynamical system all primary constraints always equal zero. This 
allows us to introduce the corresponding zero-surface (or shell) of primary constraints $S_p$. For 
the metric gravity we have $d$ primary constraints $\phi^{0\lambda}$ (see above). By following 
\cite{K&K} let us consider the case when all chains of gauge transformations are of length two, i.e., 
the Castellani generators are the linear combination of the two $C^{\lambda}_{1}(x)$ and 
$C^{\lambda}_{0}(x)$ functions, where  
\begin{eqnarray}
 C^{\lambda}_{1}(x) = \phi^{0\lambda} \; \; \; and \; \; \; C^{\lambda}_{0}(x) = [ H_{t}, 
 \phi^{0\lambda} ] + \int A^{\lambda}_{\mu}(x,y) \phi^{0\mu}(y) d^{3}y \; \; , \; \label{Cast1}
\end{eqnarray}
where $\lambda = 0, 1, \ldots, d - 1$ is the index of the chain (or index of the generating primary 
constraint), while the lower index $k$ is used to numerate all gauge generators in one chain. The 
$A^{\lambda}_{\mu}(x,y)$ are the functions which are chosen from the fact that these chains of 
generators must be finished on the surface of primary constraints $S_p$. This leads to the following 
condition for the Poisson bracket 
\begin{eqnarray}
 [ C^{\lambda}_{0}(x), H_t ] = \sum_{\nu} L^{\lambda}_{\nu}(x) \phi^{0\nu}(x) = {\rm linear} \; 
 \; {\rm combination} \; \; {\rm of} \; \; {\rm primary} \; \; {\rm constraints} \; , 
 \; \label{Cast2}
\end{eqnarray}
where $L^{\lambda}_{\nu}(x)$ are some continuous functions. Now, the Castellani generators take the 
following general form $G(\xi_{\lambda}) = \xi_{\lambda} C^{\lambda}_{0}(x) + \xi_{\lambda;0} 
C^{\lambda}_{1}(x)$, where $\xi_{\lambda}$ is the $\lambda-$th gauge parameter, while $\xi_{\lambda;0}$ 
is ts time-derivative of the first-order. The gauge parameter is a function of the spatial coordinates 
and time only, but it cannot depend upon the field itself, or upon any component of the metric 
gravitational field in our case. Moreover, in applications to the metric GR such gauge parameters can 
be used only in a completely covariant form. Segregation of some `selected' components of these 
parameters is strictly prohibited, since it devaluates the original Castellani procedure and leads to 
the results which are fundamentally wrong.  

Now, by using the criterion, Eq. (\ref{Cast2}), we obtain the following equation
\begin{eqnarray}
 [ C^{\lambda}_{0}(x), H_t ] = - [ \chi^{0\lambda}, H_{t} ] + \int [ A^{\lambda}_{\mu}(x,y), H_{t} ] 
 \phi^{0\mu}(y) d^{3}y + \int A^{\lambda}_{\mu}(x,y) [ \phi^{0\mu}(y), H_{t} ] d^{3}y \; \; , \; 
 \label{Cast3}
\end{eqnarray}
which can be used to determine the unknown function $A^{\lambda}_{\mu}(x,y)$. Formally, the second term 
in the right-hand side of this equation is already written as the linear combination of the primary 
constraints only. On the surface of primary constraints $S_p$ this term vanishes. For now this (second) 
term can be neglected. This allows us to derive the following explicit formula for the 
$C^{\lambda}_{0}(x)$ function \cite{K&K}
\begin{eqnarray}
 &&C^{\lambda}_{0}(x) = - \chi^{0\lambda} - \Bigl( \frac12 g_{00,0} g^{0\lambda} + g_{0m,0} 
 g^{\lambda m} - \frac12 g^{\lambda m} g_{00,m} \Bigr) \phi^{00} - \delta^{\lambda}_{0} 
 \phi^{0 k}_{,k} - \Bigl( \frac{2}{\sqrt{- g}} I_{nm pk} \pi^{m n} \frac{g^{\lambda p}}{g^{00}} 
 \nonumber \\
 &&- I_{mk pq} g_{\alpha\beta,l} A^{(pq)0 \alpha\beta l} \frac{g^{\lambda m}}{g^{00}} \Bigr) 
 \phi^{0 k} - \Bigl[ g^{0\lambda} g_{00,k} + 2 g^{n\lambda} g_{0n,k} + g^{n\lambda} 
 \frac{g^{0 m}}{g^{00}} ( g_{m n,k} + g_{k m,n} - g_{k n,m} ) \Bigr] \phi^{0 k} \nonumber 
\end{eqnarray} 
The Castellani generator is now determined by the relation $G(\xi_{\lambda}) = \xi_{\lambda} 
C^{\lambda}_{0}(x) + \xi_{\lambda;0} \phi^{0\lambda}(x)$ mentioned above. This generator can be 
applied to obtain the transformation of the metric tensor $g_{\alpha\beta}$, i.e., 
\begin{eqnarray}
 \delta g_{\alpha\beta} = [ G(\xi_{\lambda}), g_{\alpha\beta} ] = [ \xi_{\lambda} 
 C^{\lambda}_{0} + \xi_{\lambda;0} \phi^{0\lambda}, g_{\alpha\beta}] \; \; , \; \label{Cast4}
\end{eqnarray} 
The following transformations (see, Section 5 in \cite{K&K}) directly and unambiguously lead to 
the diffeomorphism invariance, which is the well known gauge invariance of the free gravitational 
field(s). Briefly, we can say that the Einstein equations of motion $G_{\mu\nu} = R_{\mu\nu} - 
\frac12 g_{\mu\nu} R = 0$ are transformed into a linear combination of the same equations. Note 
that such an invariance is true only on-shell of the primary constraints (see, discussions in 
\cite{K&K} and \cite{Petr}). Formally, the Dirac closure of the metric GR is needed to be checked 
on the same zero-surface of primary constraints only \cite{K&K}.  

Currently, there are only two known Hamiltonian formulations \cite{Dir58} and \cite{K&K} developed 
for the metric GR which reproduce the actual diffeomorphism invariance directly and transparently. 
In contrast with this, numerous Hamiltonian formulations of metric GR based on the ADM dynamical 
variables (see, Appendix B) fail at this crucial point. Note that for all approaches which are 
directly based on the $\Gamma - \Gamma$ Lagrangian of the metric GR, such a reconstruction of the 
diffeomorphism invariance is a relatively simple problem (see, e.g., \cite{Saman}). In contrast with 
this, for any Hamiltonian-based formulation the complete solution of similar problem requires a 
substantial work. On the other hand, analytical derivation of the diffeomorphism invariance is a very 
good test for the total $H_t$ and canonical $H_C$ Hamiltonians as well as for all primary 
$\phi^{0\sigma}$ and secondary $\chi^{0\sigma}$ constraints derived in any new Hamiltonian formulation 
of the metric GR. Any mistake made either in the $H_t, H_C$ Hamiltonians, or in the $\phi^{0\lambda}$ 
and $\chi^{0\sigma}$ constraints leads to the loss of true diffeomorphism invariance for the free 
gravitational field.   
   
\subsection{Hamilton equations of motion for the free gravitational field}

In general, if we know the total $H_t$ Hamiltonian, Eq. (\ref{eq1}), then we can derive the Hamilton 
equations of motion which describe the time-evolution of all essential dynamical variables in the 
metric GR, i.e., time-evolution of each component of the metric tensor $g_{\alpha\beta}$ and momentum 
tensor $\pi^{\gamma\rho}$. These equations are \cite{Fro1} 
\begin{eqnarray}
 \frac{d g_{\alpha\beta}}{d x_0} = [ g_{\alpha\beta}, H_{t} ] \; \; \; {\rm and} \; \; \; 
 \frac{d \pi^{\gamma\rho}}{d x_0} = [ \pi^{\gamma\rho}, H_{t} ] \; , \; \label{eq20}
\end{eqnarray}
where the notation $x_0$ denotes the temporal variable. In particular, for the spatial components 
$g_{ij}$ of the metric tensor $g_{\alpha\beta}$ one finds the following equations
\begin{eqnarray}
 \frac{d g_{ij}}{d x_0} &=& [ g_{ij}, H_{t} ] = [ g_{ij}, H_{c} ] = \frac{2}{\sqrt{-g} g^{00}} 
 I_{(ij)pq} \pi^{pq} - \frac{1}{g^{00}} I_{(ij)pq} B^{(p q 0|\mu \nu k)} g_{\mu\nu,k} \; \label{eq25} \\
 &=& \frac{2}{\sqrt{-g} g^{00}} I_{(ij)pq} \Bigl[ \pi^{pq} - \frac12 \sqrt{-g} B^{(p q 0|\mu \nu k)} 
 g_{\mu\nu,k} \Bigr] \; , \nonumber 
\end{eqnarray}
where the notation $I_{(ij)pq}$ stands for the $(ij)-$symmetrized values of the $I_{ijpq}$ tensor 
defined in Eq.(\ref{I}), i.e., 
\begin{equation}
 I_{(ij)pq} = \frac12 \Bigl( I_{ijpq} + I_{jipq} \Bigr) = \frac{1}{d - 2} g_{ij} g_{pq} - \frac12 
 ( g_{ip} g_{jq} + g_{iq} g_{jp} ) \; \; \; .
\end{equation}
Analogously, for the $g_{0\sigma}$ components of the metric tensor one finds the following equations 
of time-evolution
\begin{eqnarray}
 \frac{d g_{0\sigma}}{d x_0} = [ g_{0\sigma}, H_{t} ] = g_{0\sigma,0} \; , \; \; \label{eq253}
\end{eqnarray}
since all $g_{0\sigma}$ components commute with the canonical Hamiltonian $H_C$, Eq.(\ref{eq5}), while 
all $g_{ij}$ commute with the primary constraints $\phi^{0\sigma}$. This result could be expected, 
since the equation, Eq.(\ref{eq253}), is, in fact, a definition of the $\sigma-$velocities (or 
$g_{0\sigma,0}$-velocities), where $\sigma = 0, 1, \ldots, d - 1$.   

The Hamilton equations for the tensor components of momentum $\pi^{\alpha\beta}$, Eq.(\ref{eq20}), are 
substantially more complicated. They are derived by calculating the Poisson brackets between each term 
in $H_{t}$ and $\pi^{\gamma\rho}$. This general formula takes the form
\begin{eqnarray}
 \frac{d \pi^{\alpha\beta}}{d x_0} &=& - [ H_{t}, \pi^{\alpha\beta} ] = - \Bigl[ 
 \frac{I_{mnpq}}{\sqrt{-g} g^{00}}, \pi^{\alpha\beta} \Bigr] \pi^{mn} \pi^{pq} \nonumber \\
 &+& \Bigl[ \frac{I_{mnpq}}{g^{00}}, \pi^{\alpha\beta} \Bigr] \pi^{mn} B^{(p q 0|\mu \nu k)} 
 g_{\mu\nu,k} + \frac{1}{g^{00}} I_{mnpq} \pi^{mn}\Bigl[ B^{(p q 0|\mu \nu k)}, \pi^{\alpha\beta} 
 \Bigr] g_{\mu\nu,k} + \ldots \; . \label{eq255}
\end{eqnarray}
Let us determine the first Poisson bracket in this formula (other terms in Eq.(\ref{eq255}) are considered 
analogously, i.e., term-by-term). The explicit expression for this term is 
\begin{eqnarray}
 &-& \Bigl[ \frac{I_{mnpq}}{\sqrt{-g} g^{00}}, \pi^{\alpha\beta} \Bigr] \pi^{mn} \pi^{pq} = - \frac{[ 
 I_{mnpq}, \pi^{\alpha\beta}]}{\sqrt{-g} g^{00}} \pi^{mn} \pi^{pq} - [ \frac{1}{\sqrt{-g} g^{00}}, 
 \pi^{\alpha\beta} \Bigr] I_{mnpq} \pi^{mn} \pi^{pq} \; \; . \label{eq256}
\end{eqnarray}
Thus, we have the three following cases: (1) for a pair of space-like indexes, i.e., for $(\alpha\beta) = 
(a b)$, where one finds  
\begin{eqnarray}
 \Bigl( \frac{d \pi^{a b}}{d x_0}\Bigr)_1 = -\frac{2}{d - 2} g_{m n} \pi^{m n} \pi^{a b} + 2 g_{m p} 
 \pi^{m a} \pi^{p b} + \frac{I_{mnpq}}{2 \sqrt{-g} g^{00}} g^{a b} \pi^{m n} \pi^{p q} \; \; , \; 
 \label{eq257}
\end{eqnarray}
while for the mixed pair of indexes $(\alpha\beta) = (0 a)$ the analogous expression is 
\begin{eqnarray}
 \Bigl( \frac{d \pi^{0 a}}{d x_0}\Bigr)_1 = \frac{I_{mnpq}}{2 \sqrt{-g} g^{00}} g^{0 a} 
 \pi^{m n} \pi^{p q} \; . \; \label{eq2561}
\end{eqnarray}
Finally, for the temporal pair of indexes $(\alpha\beta) = (0 0)$ pair one finds
\begin{eqnarray}
 \Bigl( \frac{d \pi^{0 0}}{d x_0}\Bigr)_1 = \frac{I_{mnpq}}{2 \sqrt{-g}} \Bigl( 1 + 
 \frac{2}{(g^{00})^{2}} \Bigr) \pi^{mn} \pi^{pq} \; . \; \label{eq2562}
\end{eqnarray}
In general, analytical calculations of other Poisson brackets in the formula, Eq. (\ref{eq255}), is a 
straightforward task, but the final formula contains more than 150 terms. This drastically complicates 
all operations with the formula, Eq. (\ref{eq255}), for the $\frac{d \pi^{\gamma\rho}}{d x_0}$ 
derivative. Actual analytical and numerical computations of the time-evolution of the free 
gravitational field(s) can be performed by using modern packages of computer algebra. Nevertheless, the 
complete Hamilton equations of motion for the free gravitational field(s) in metric GR have been derived 
and written explicitly in a closed analytical forms.  

\subsection{On the general form of canonical transformations in the metric GR}

As is well known all canonical transformations for an arbitrary Hamiltonian system form a closed algebraic 
group. This means that in any Hamiltonian system: (1) consequence of the two canonical transformations is 
the new canonical transformation, (2) identical transformation of dynamical variables is the canonical 
transformation, (3) any canonical transformation has its inverse transformation which is also canonical 
and unique. In general, there are quite a few canonical transformations in the metric General Relativity, 
and some of them can be used to simplify either Hamiltonian(s), or secondary constraints, or some other 
crucial quantities, including a few important Poisson brackets. As is well known (see, e.g., \cite{LLTF}, 
\cite{Carm}) the metric General Relativity is a non-linear theory which cannot rigorously be linearized 
even in second-order approximation. Therefore, the linear canonical transformations of dynamical variables 
are no interest for the Hamiltonian formulations which have been developed for the metric GR. Furthermore, 
it can be shown that among all possible non-linear canonical transformations the following `special' 
transformations play a great role in derivation of all new Hamiltonian formulations of the metric GR. 
These special canonical transformations can be written in the form: $\{ g_{\alpha\beta}, \pi^{\mu\nu}\} 
\rightarrow \{ g_{\alpha\beta}, \Pi^{\rho\sigma}\}$, where the new momenta $\Pi^{\rho\sigma}$ are the 
linear (or quasi-linear) combinations of old momenta $\pi^{\mu\nu}$ and all spatial derivatives of the 
components of metric tensor. In general, such a combination is written in the form 
\begin{eqnarray}
 \Pi^{\gamma\rho} = A^{\gamma\rho}_{\lambda\sigma} \pi^{\lambda\sigma} - \frac12 \sqrt{-g} 
 C^{\gamma\rho}_{\lambda\sigma} D^{\lambda\sigma 0 \mu\nu k} g_{\mu\nu, k} \; \; , \; \label{cantr}
\end{eqnarray}
where $A^{\gamma\rho}_{\lambda\sigma}$ and $C^{\gamma\rho}_{\lambda\sigma}$ are the constant (or numerical) 
tensors, while the tensor-functions $D^{\gamma\rho 0 \mu\nu k}$ are the cubic polynomial of all 
contravariant components of the metric tensor $g^{\alpha\beta}$. When the both 
$A^{\gamma\rho}_{\lambda\sigma}$ and $C^{\gamma\rho}_{\lambda\sigma}$ tensors are the diadas of the two 
substitution tensors, i.e., each of them equals to the product $\delta^{\gamma}_{\lambda} 
\delta^{\rho}_{\sigma}$, then from Eq.(\ref{cantr}) one finds 
\begin{eqnarray}
 \Pi^{\gamma\rho} = \pi^{\gamma\rho} - \frac12 \sqrt{-g} D^{\gamma\rho 0 \mu\nu k} g_{\mu\nu, k} 
 \; \; . \; \label{cantr1}
\end{eqnarray}

At this moment all known canonical transformations which arise in the metric GR are represented in the 
form of Eq.(\ref{cantr}), or Eq.(\ref{cantr1}). In particular, the canonical transformation which 
relates the two correct Hamiltonian formulations currently known in the metric GR, i.e., Hamiltonian 
formulation developed by Dirac \cite{Dir58} and K$\&$K \cite{K&K} Hamiltonian formulation, is 
represented in the form of Eq.(\ref{cantr1}). Our new canonical transformation of dynamical variables 
of metric GR, which is described below, is also written in the form of Eq.(\ref{cantr1}). Furthermore, 
it can be shown that transformations of the dynamical variables of metric GR, chosen in the form of
Eq.(\ref{cantr1}), will preserve the complete diffeomorphism as a gauge symmetry of the free 
gravitational field. It is clear that such `special' form of canonical transformations in the metric 
General relativity is substantially determined by the $\Gamma - \Gamma$ Lagrangian presented in Section 
II. Indeed, the $\Gamma - \Gamma$ Lagrangian, Eq.(\ref{eq05}), is a polynomial of power three upon the 
contravariant $g^{\alpha\beta}$ components of metric tensor (i.e., it is a cubic polynomial) and a 
quadratic function of the spatial velocities $g_{mn,0}$. Alternative transformations of dynamical 
variables of the metric GR, which cannot be written in the form of Eq.(\ref{cantr}) and/or 
Eq.(\ref{cantr1}), probably, are not canonical, and they cannot be used to relate the two different 
(but canonical) Hamiltonian formulations of metric GR. 

In order to understand a special role of the canonical transformations chosen in the form of 
Eq.(\ref{cantr1}), consider the difference between the two following differential forms 
\begin{eqnarray}
 \pi^{\alpha\beta} dg_{\alpha\beta} - \Pi^{\alpha\beta} dg_{\alpha\beta} = \frac12 \sqrt{-g}
  D^{\alpha\beta 0 \mu\nu k} g_{\mu\nu, k} dg_{\alpha\beta} \; \; . \; \label{cantr2}
\end{eqnarray} 
This form is, in fact, the difference between the two relative integral invariants, or Poincar\'{e} 
integral invariants. The well known Poincar\'{e} theorem states that the transformation $\{ 
g_{\alpha\beta}, \pi^{\mu\nu}\} \rightarrow \{ g_{\alpha\beta}, \Pi^{\rho\sigma}\}$ of dynamical 
variables will be canonical if (and only if) such a difference of the two relative integral invariants, 
i.e., the expression on the right side of the last equation, will be a total differential. This means 
that canonicity of the new variables will be obeyed only in those cases, when the expression $\frac12 
\sqrt{- g} D^{\alpha\beta 0 \mu\nu k} g_{\mu\nu, k} dg_{\alpha\beta}$ is the total differential of 
some function. Now, the integrability conditions are written in the following forms 
\begin{eqnarray}
  \frac{\partial \Bigl[\sqrt{-g} D^{\alpha\beta 0 \mu\nu k} \Bigr]}{\partial g_{\lambda\sigma}} = 
  \frac{\partial \Bigl[\sqrt{-g} D^{\lambda\sigma 0 \mu\nu k} \Bigr]}{\partial g_{\alpha\beta}} 
  \; \; , \; \label{Integrab1}
\end{eqnarray}  
which are completely equivalent to the conditions $[ \Pi^{\alpha\beta}, \Pi^{\lambda\sigma}] = 0$ for 
the fundamental Poisson brackets of the new dynamical variables $\{ g_{\alpha\beta}, 
\Pi^{\lambda\sigma}\}$. Indeed, for these Poisson brackets we can write 
\begin{eqnarray}
 0 = [ \Pi^{\alpha\beta}, \Pi^{\lambda\sigma}] = - \frac12 [ \pi^{\alpha\beta}, \sqrt{-g} 
 D^{\lambda\sigma 0 \mu\nu k}] g_{\mu\nu, k} + \frac12 [ \pi^{\lambda\sigma}, \sqrt{-g} 
 D^{\alpha\beta 0 \mu\nu k}] g_{\mu\nu, k} \; \; , \; \label{Integrab2}
\end{eqnarray}   
which are easily reduced to the form of Eq.(\ref{Integrab1}). It is interesting to compare this 
equation to Eq. (29) from \cite{FK&K} which also has the form of Poisson brackets on the one hand, 
as well as integrability conditions for some differential 1-form on the other. This 1-form is the  
Poincar\'{e} integral invariant. As follows from this discussion all fundamental Poisson brackets 
between basic dynamical variables of an arbitrary Hamiltonian system can be split into three 
different groups and two of these groups can be considered as the systems of integrability 
conditions. In particular, the Poisson brackets from the third group, i.e. $[ p_i, p_j ] = 0$, 
are, in fact, the integrability conditions for some first-order differential form \cite{Fland} 
which is written in the coordinate space. Analogously, the these Poisson brackets from the second 
group, i.e., $[ q_i, q_j ] = 0$, can be considered as the integrability conditions for same 
first-order differential form written in the momentum representation. 

\section{Reduction of the canonical Hamiltonian to its natural form}

In this Section we reduce the canonical Hamiltonian $H_C$ to its natural form, which will play 
a significant role in numerous applications to the metric gravity. We perform such a reduction 
of $H_C$ by using some new canonical transformation of the dynamical variables $g_{\alpha\beta}$ 
and $\pi^{\rho\sigma}$ defined above. First, let us write the canonical Hamiltonian, 
Eq.(\ref{eq5}), in the form
\begin{eqnarray}
 H_C &=& \frac{I_{mnpq}}{\sqrt{-g} g^{00}} \Bigl[ \pi^{mn} \pi^{pq} - \sqrt{-g} \pi^{mn} 
 B^{(p q 0|\mu \nu k)} g_{\mu\nu,k} + \frac14 (- g) B^{(m n 0|\mu \nu k)} 
 B^{(p q 0|\alpha \beta l)} g_{\mu\nu,k} g_{\alpha\beta,l} \Bigr] \nonumber \\
 &+& \frac14 \sqrt{-g} \Bigl \{ \frac{1}{g^{00}} I_{mnpq} B^{([mn] 0|\mu\nu k)} 
 B^{(p q 0|\alpha \beta l)} - B^{\mu\nu k \alpha\beta l}\Bigr \} g_{\mu\nu,k} 
 g_{\alpha\beta,l} \; , \; \label{eq5a} 
\end{eqnarray}
which is more appropriate for our purposes in this study. In Eq. (\ref{eq5a}) the notation 
$B^{([mn] 0|\mu\nu k)}$ stands for the $B^{(m n 0 \mid \mu\nu k)}$ cubic function of the 
contravariant components of the metric tensor which is completely anti-symmetric in respect 
to all permutations of the $m$ and $n$ indexes. The explicit formula for the 
$B^{([mn] 0|\mu\nu k)}$ function is 
\begin{eqnarray}
 B^{([mn] 0|\mu\nu k)} &=& g^{m k} g^{n \nu} g^{\nu 0} - g^{n k} g^{m \nu} g^{\nu 0} 
 + \frac12 \Bigl( g^{n \mu} g^{m \nu} g^{k 0} + g^{n k} g^{\mu \nu} g^{m 0} - 
 g^{m \mu} g^{n \nu} g^{k 0} \nonumber \\ 
 &-& g^{m k} g^{\mu \nu} g^{m 0} \Bigr) \; . \; \label{AsBcoef}
\end{eqnarray}
Now, we can see that the first term in $\Bigl[ \ldots \Bigr]$ brackets in Eq. (\ref{eq5a}) 
can be written as a pure quadratic function of the new $P^{mn} = \pi^{mn} - \frac12 \sqrt{-g} 
B^{(m n 0|\mu\nu k)} g_{\mu\nu, k}$ variables (spatial momenta), i.e., 
\begin{eqnarray}
 H_C &=& \frac{I_{mnpq}}{\sqrt{-g} g^{00}} \Bigl( \pi^{mn} - \frac12 \sqrt{-g} 
 B^{(m n 0|\mu\nu k)} g_{\mu\nu, k} \Bigr) \Bigl( \pi^{pq} - \frac12 \sqrt{-g} 
 B^{(p q 0|\alpha\beta l)} g_{\alpha\beta, l} \Bigr) \nonumber \\
 &+& \frac14 \sqrt{-g} \Bigl\{ \frac{1}{g^{00}} I_{mnpq} B^{([mn] 0|\mu\nu k)} 
 B^{(p q 0|\alpha \beta l)} - B^{\mu\nu k \alpha\beta l}\Bigr\} g_{\mu\nu,k} 
 g_{\alpha\beta,l} + T_1 + T_2 \; \; , \; \label{H_Cnew}
\end{eqnarray}
where the two additional terms $T_1$ and $T_2$ are:  
\begin{eqnarray}
 T_1 = \frac{I_{mnpq}}{2 \sqrt{-g} g^{00}} [ \pi^{mn}, \sqrt{- g}] B^{(p q 0|\alpha \beta l)} 
 g_{\alpha\beta,l} = - \frac{I_{mnpq} g^{mn}}{2 g^{00}} B^{(p q 0|\alpha \beta l)} 
 g_{\alpha\beta,l} \; \; \; \label{eq5b} 
\end{eqnarray}
and 
\begin{eqnarray}
 T_2 &=& - \frac{I_{mnpq}}{2 g^{00}} [ B^{(m n 0|\mu \nu k)}, \pi^{pq} ] g_{\mu\nu,k} = 
 - \frac{I_{mnpq}}{2 g^{00}} \Bigl[ \frac12 \Bigl( g^{\mu p} g^{m q} + g^{\mu q} g^{m p} 
 \Bigr) g^{n \nu} g^{k 0} \nonumber \\
 &+& \frac12 g^{\mu m} \Bigl( g^{n p} g^{\nu q} + g^{n q} g^{\nu p} \Bigr) g^{k 0} 
 + \frac12 g^{\mu m} g^{n \nu} \Bigl( g^{k p} g^{0 q} + g^{k q} g^{0 p} \Bigr) \nonumber \\
 &-& \frac12 \Bigl( g^{m p} g^{n q} + g^{m p} g^{n q} \Bigr) g^{k 0} g^{\mu\nu} - 
 \frac12 g^{m n} \Bigl( g^{p k} g^{q 0} + g^{p 0} g^{q k} \Bigr) - \frac12 g^{m n} 
 g^{k 0} \Bigl( g^{\mu p} g^{\nu q} + g^{\mu q} g^{\nu p} \Bigr) \nonumber \\ 
 &-& \Bigl( g^{m p} g^{k q} + g^{m q} g^{k p} \Bigr) g^{n \nu} g^{\mu 0} - g^{m k} 
 \Bigl( g^{n p} g^{\nu q} + g^{n q} g^{\nu p} \Bigr) g^{\mu 0} - \frac12 g^{m k} 
 g^{n \nu} \Bigl( g^{\mu p} g^{0 q} + g^{\mu q} g^{0 p} \Bigr)  \nonumber \\
 &+& \frac12 \Bigl( g^{m p} g^{n q} + g^{m q} g^{n p} \Bigr) g^{\nu k} g^{0 \mu} + 
 \frac12 g^{m n} \Bigl( g^{\nu p} g^{k q} + g^{\nu q} g^{k p} \Bigr) g^{\mu 0} + 
 \frac12 g^{m n} g^{\nu k} \Bigl( g^{p 0} g^{\mu q} + g^{0 q} g^{\mu p} \Bigr) \nonumber \\
 &+& \frac12 \Bigl( g^{k p} g^{m q} + g^{k q} g^{m p} \Bigr) g^{\nu k} g^{0 \mu} 
 + \frac12 g^{k m} \Bigl( g^{\mu p} g^{\nu q} + g^{p \nu} g^{\mu q} \Bigr) g^{n 0} 
 + \frac12 g^{k m} g^{\mu \nu} \Bigl( g^{n p} g^{0 q} \nonumber \\
 &+& g^{n q} g^{0 p} \Bigr) \Bigr] g_{\mu\nu,k} \; \; , \; \label{eq5c} 
\end{eqnarray}
respectively. 

Now, we can explicitly introduce the new momenta $P^{\gamma\rho}$ which are written in the 
following form 
\begin{eqnarray}
 P^{\gamma\rho} = \pi^{\gamma\rho} - \frac12 \sqrt{-g} B^{(\gamma\rho 0|\mu\nu k)} 
 g_{\mu\nu, k} \; \; , \; \label{canvar}
\end{eqnarray}
where $\pi^{\gamma\rho}$ are the `old' momenta used in \cite{K&K}. These new momenta can be 
considered as the contravariant components of the tensor of one `united' momentum $P$ of the 
metric gravitational field. Note that the explicit expressions for the old velocities written 
in terms of new momenta $P^{ab}$ are even simpler $g_{mn, 0} = \frac{1}{\sqrt{-g} g^{00}} 
I_{m n q p} P^{pq}$, than the expression given by Eq.(\ref{veloc}). The explicit formulas for 
the primary constraints are also simpler: $P^{0\gamma} \approx 0$ for $\gamma = 0, 1, \ldots, 
d - 1$. The generalized coordinates are chosen in the old (or traditional) form, i.e., they 
coincide with the covariant components of the metric tensor $g_{\alpha\beta}$. Such a choice 
of the generalized coordinates provides a number of additional advantages in applications to 
the metric GR. For instance, by using the metric tensor one can rise and lower indexes in 
arbitrary vectors and tensors. Also, all covariant and contravariant derivatives of the metric 
tensor always equal zero, i.e., this tensor behaves as a constant during these operations. 
More unique and remarkable properties of the metric tensor are discussed, e.g., in 
\cite{Kochin}. For the purposes of this study it is important to note that our new system of 
dynamical variables contains the same coordinates $g_{\alpha\beta}$ and new momenta 
$P^{\gamma\rho}$. 

The Poisson brackets between our new dynamical variables can easily be determined by using the 
known values of Poisson brackets written in the old dynamical variables $\Bigl\{ g_{\alpha\beta}, 
\pi^{\gamma\rho} \Bigr\}$ defined above. Indeed, for the corresponding Poisson brackets one finds: 
$[ g_{\alpha\beta}, P^{\gamma\rho} ] = [ g_{\alpha\beta}, \pi^{\gamma\rho} ] = 
\Delta^{\gamma\rho}_{\alpha\beta} = \frac12 \Bigl( \delta^{\gamma}_{\alpha} \delta^{\sigma}_{\beta} 
+ \delta^{\sigma}_{\alpha} \delta^{\gamma}_{\beta} \Bigr), [ g_{\alpha\beta}, g_{\gamma\rho} ] = 0$ 
and $[ P^{\alpha\beta}, P^{\gamma\rho} ] = 0$. The last equality we consider in detail
\begin{eqnarray}
 & &[ P^{\alpha\beta}, P^{\gamma\rho} ] = [ \pi^{\alpha\beta}, \pi^{\gamma\rho} ] - \frac12 [ 
 \sqrt{-g} B^{(\alpha\beta 0|\mu\nu k)}, \pi^{\gamma\rho} ] g_{\mu\nu, k} + \frac12 [ \sqrt{-g} 
 B^{(\alpha\beta 0|\lambda\sigma, l)}, \pi^{\gamma\rho} ] g_{\lambda\sigma, l} \nonumber \\
 &+& [ \sqrt{-g} B^{(\alpha\beta 0|\mu\nu k)} g_{\mu\nu, k}, \sqrt{-g} B^{(\gamma\rho 0|\lambda\sigma l)} 
 g_{\lambda\sigma, l} ] \; , \; \label{PBV}
\end{eqnarray}
where the first and last terms equal zero identically, since the variables $g_{\alpha\beta}$ and 
$\pi^{\mu\nu}$ are canonical. This directly leads to the formula 
\begin{eqnarray}
 [ P^{\alpha\beta}, P^{\gamma\rho} ] = - \frac12 [ \sqrt{-g} B^{(\alpha\beta 0|\mu\nu k)}, 
 \pi^{\gamma\rho} ] g_{\mu\nu, k} + \frac12 [ \sqrt{-g} B^{(\alpha\beta 0|\lambda\sigma, l)}, 
 \pi^{\gamma\rho} ] g_{\lambda\sigma, l} \; \; . \label{PBV1}
\end{eqnarray}
Now, we can replace the dummy indexes in the second term of this equation by the values which coincide 
with the corresponding dummy indexes in the first term, i.e., $\lambda \rightarrow \mu, \sigma 
\rightarrow \nu$ and $l \rightarrow k$. This substitution reduces Eq. (\ref{PBV1}) to the form (compare 
with Eq.(\ref{Integrab2}) from above)
\begin{eqnarray}
 [ P^{\alpha\beta}, P^{\gamma\rho} ] = - \frac12 [ \sqrt{-g} B^{(\alpha\beta 0|\mu\nu k)}, \pi^{\gamma\rho} ] 
 g_{\mu\nu, k} + \frac12 [ \sqrt{-g} B^{(\alpha\beta 0|\mu\nu, k)}, \pi^{\gamma\rho} ] g_{\mu\nu, k} = 0 
 \; , \; \label{PBV2}
\end{eqnarray}
since it is the difference of the two identical expressions. This shows that the new dynamical variables 
$\{ g_{\alpha\beta}, P^{\mu\nu}\}$ are also canonical, and they can be used in the metric gravity, since they 
are canonically related to the old set of K$\&$K variables $\{ g_{\alpha\beta}, \pi^{\mu\nu}\}$ \cite{K&K}. 

As follows from the formulas derived above the canonical Hamiltonian $H_C$ is reduced to the following final 
form 
\begin{eqnarray}
 H_C &=& \frac{I_{mnpq}}{\sqrt{-g} g^{00}} P^{mn} P^{pq} + \frac14 \sqrt{-g} \Bigl[ \frac{I_{mnpq}}{g^{00}} 
 B^{([mn] 0|\mu\nu k)} B^{(p q 0|\alpha \beta l)} - B^{\mu\nu k \alpha\beta l}\Bigr] g_{\mu\nu,k} 
 g_{\alpha\beta,l} \nonumber \\ 
 &-& \frac{I_{mnpq}}{2 g^{00}} g^{mn} B^{(pq 0| \alpha\beta l)} g_{\alpha\beta,l} + T_2 \; , \; \label{eq5d} 
\end{eqnarray}
which can be re-written in the following symbolic form 
\begin{eqnarray}
  H_C = \frac12 \sum^{n}_{i,j=1} \hat{M}_{ij}(q_1, q_2, \ldots, q_n) p_i p_j + \sum^{n}_{i,j=1} 
  \hat{V}_{mn}(q_1, q_2, \ldots, q_n) \; , \; \label{ClassH} 
\end{eqnarray}
where $\hat{M}$ is a positively defined and invertable $n \times n$ (symmetric) matrix which is often called the matrix of inverse masses. 
The $\hat{V}$ matrix in this equation is an arbitrary, in principle, symmetric $n \times n$ matrix which is called the potential matrix 
(or matrix of the potential energy). Here $n$ is the total number of generalized coordinates $q_1, q_2, \ldots, q_n$. Each matrix element 
of the potential matrix $\hat{V}$ in Eq. (\ref{ClassH}) is a polynomial which depends upon these generalized coordinates. Also, in 
Eq. (\ref{ClassH}) the notations $p_i$ and $p_j$ designate the momenta conjugate to the corresponding generalized coordinates $q_i$ and $q_j$, 
respectively, i.e., $[ q_k, p_l] = \delta_{kl}$. In classical mechanics the phase space is flat, and, therefore, the both covariant and 
contravariant components of any vector coincide with each other. The form of the Hamiltonian $H_C$, Eq. (\ref{ClassH}), is called normal, and 
it is well known in classical mechanics of Hamiltonian systems. Furthermore, more than 90 \% of all problems ever solved in classical 
Hamiltonian mechanics have Hamiltonians which are already written in the normal form, or their Hamiltonians can easily be reduced to their 
normal forms by some canonical transformation(s) of variables. The idea of reducing the Hamiltonians to their normal forms goes back to 
Poincar\'{e} \cite{Poin}. A separate area of modern mathematical physics is the study of the normal forms of different Hamiltonians in the 
vicinities of equilibrium positions \cite{Birk} (see also discussion and references in the Appendix 7 from \cite{Arnold}). 
 
To improve the overall quality of our analogy between metric GR and classical Hamiltonian mechanics let us introduce the new set of dynamical 
variables which include the total momentum of the free gravitational field $P = g_{\alpha\beta} P^{\alpha\beta}$ (it is a tensor invariant) 
and its tensor `projections' $P_{\alpha}^{\beta} = g_{\alpha\gamma} P^{\gamma\beta}$. The corresponding space-like quantities $P = g_{mn} 
P^{mn}$ and $P_{m}^{n} = g_{m p} P^{p n}$ are included in our canonical Hamiltonian $H_C$, Eq. (\ref{eq5d}). By using our formulas presented 
above one easily finds a few following Poisson brackets:
\begin{eqnarray}  
 &[& P, P^{ab} ] = [ g_{mn}, P^{ab} ] P^{mn} = \Delta^{ab}_{mn} P^{mn} = P^{ab} \; , \; [ g_{cd}, P ] = g_{mn} [ g_{cd}, P^{mn} ] 
 = g_{cd} \nonumber \\
 &[& g_{\alpha\beta}, P^{\gamma}_{\sigma} ] = \frac12 ( g_{\beta\sigma} \delta^{\gamma}_{\alpha} + g_{\alpha\sigma} 
 \delta^{\gamma}_{\beta} ) \; \; , \;  [ g^{\alpha\beta}, P ] = g^{\alpha\beta} \; \; \; 
 \nonumber 
\end{eqnarray}
and others. By using the total momentum $P$ and its tensor projections (i.e., $P^{\alpha\beta}, P^{\gamma}_{\sigma}$, etc) one can write the 
Hamilton equations in the form which almost coincides with analogous equations known for Hamiltonian systems in classical mechanics. This is 
another interesting direction for future development of the Hamiltonian formulation(s) of metric GR. Other applications of our new canonical 
variables $\{ g_{\lambda\kappa}, P^{\alpha\beta} \}$ to some interesting problems in metric GR  will be considered elsewhere. Relations 
between our dynamical variables $\{ g_{\lambda\kappa}, P^{\alpha\beta} \}$ and analogous variables used in Dirac formulation of the metric 
General Relativity $\{ g_{\lambda\kappa}, p^{\alpha\beta} \}$ are discussed in the Appendix A. 

\section{Jacobi equation for the free Gravitational Field in our Hamiltonian formulation}

In general, if we have found the canonical Hamiltonian $H_C$ and properly defined all essential momenta $\pi^{mn}$ of the free gravitational field, 
then it is possible to derive the famous Jacobi equation which governs propagation and time-evolution of the free gravitational field. For the first 
time the Jacobi equation for the free gravitational field has been derived in our earlier paper \cite{Fro1}, which was based the Hamiltonian 
approach developed in \cite{K&K}. In this study to describe the free gravitational field we apply different dynamical variables $\{ g_{\lambda\kappa}, 
P^{\alpha\beta} \}$ which are canonically related to the $\{ g_{\lambda\kappa}, \pi^{\alpha\beta} \}$ dynamical variables from \cite{K&K}. Now, we want 
to derive analogous Jacobi equation written in our new dynamical variables $\{ g_{\lambda\kappa}, P^{\alpha\beta} \}$. Complete derivation of the  
Jacobi equation(s) for the free gravitational field in the metric gravity is very complex and requires many pages of additional text. Some day it  
would be nice to write a complete review article about Jacobi equation(s) for the free gravitational field in the metric gravity, but in this study 
we have to restrict ourselves to a brief derivation of the Jacobi equation by varying the gravitational action $S$ written in the form of temporal 
integral of the $\Gamma - \Gamma$ Lagrangian $L_{\Gamma - \Gamma}$. This can be written in the form  
\begin{eqnarray}
 \delta S(g_{\alpha\beta}(t), t) = \delta \int_{\gamma} L_{\Gamma - \Gamma}(\tau, g_{\alpha\beta}(\tau), g_{\alpha\beta; 0}(\tau)) d\tau \; , 
 \; \label{Jacobi} 
\end{eqnarray}
where the integral is taken along the extremal $\gamma$ which connects the initial $t_0$-point and the final $B-$point (or $t_f = t$-point). The 
final point is a free point which is varied (variations between extremals). Note that the variation of any action, including the gravitational 
action $\delta S(g_{\alpha\beta}(t), t)$ between two extremals is always the first (total) differential of the function $S$ (or action $S$) 
\cite{GF}. Furthermore, for any Hamiltonian system the differential $dS$ is written in the form $dS = \sum_{i=1}^{n} p_i dy_i - H dt$, where $H$ 
is the Hamiltonian, while $p_i$ are the momenta. From the expression for the first differential $dS$ one finds the two following equations $H = 
-\frac{\partial S}{\partial t}$ and $p_i = \frac{\partial S}{\partial y_i}$. From these equations we find that $S$, as a function of the 
coordinates of the final point $t = t_f$, satisfies the following equation
\begin{eqnarray}
 \frac{\partial S}{\partial t} + H\Bigl( t; y_1, \ldots, y_n; \frac{\partial S}{\partial y_1}, \ldots, \frac{\partial S}{\partial y_n} \Bigr) = 0
 \; , \; \label{Jacobi1} 
\end{eqnarray}  
which is called the Jacobi equation. Here we have assumed that all extremals of our problem, which begin at the given initial point do not intersect 
each other, but form a central field of extremals \cite{GF}. In other words, our extremal is embedded in a central field of extremals which start at 
a given initial point $t_{0}$. It appears that the canonical Hamilton equations: $\frac{d q_i}{dt} = \frac{\partial H}{\partial p_i}$ and $\frac{d 
p_i}{dt} = - \frac{\partial H}{\partial q_i}$ are the system of characteristic equations for the Jacobi equation which is non-linear (see, e.g., 
\cite{Tric}, \cite{Fedor} and \cite{Cour}). This is the shortest way to the Jacobi equation from the canonical Hamilton equations. The direct physical 
sense of the Jacobi equation and its solutions is straightforward, and it was formulated in our earlier paper \cite{Fro1} as follows: {\it the real 
trajectory of the system propagates in time from the end point of one (old) Lagrange extremal to the end point located at the new Lagrange extremal, 
if it satisfies the Jacobi equation}. In other words, the Hamilton-Jacobi equation is the necessary condition for some currently known system's 
trajectory propagate in the nearest future to reach (at fixed time) the end point of the `new' extremal.

For the free gravitational field in metric GR there are a few additional complications, since the field itself is a tensor and metric gravity is a 
dynamical system with constraints. Nevertheless, it is possible to write analogous differential forms $dS = \pi^{mn} dg_{mn} - H_C dt$ and $dS = 
\pi^{\alpha\beta} dg_{\alpha\beta} - H_t dt$ in multi-dimensional phase spaces, where the first $dS = \pi^{mn} dg_{mn} - H_C dt$ form is defined in 
the $d (d - 1) + 1$-dimensional phase space, while another such a form $dS = \pi^{\alpha\beta} dg_{\alpha\beta} - H_t dt$ is defined in the $d (d + 
1) + 1$-dimensional phase space. The both these spaces are odd-dimensional and, therefore, it is correct to discuss the differential operators such 
as $curl$ (or $rotor$) defined in each of these spaces and derive the closed system of the canonical Hamilton equations (applicability of the Stoke's 
theorem is discussed, e.g., in \cite{Fland} and references therein). Then, by using this system of the canonical Hamilton equations we need to restore 
the original (non-linear) Jacobi equation for which these Hamilton equations are the system of characteristic equations.  

In our dynamical variables we can write the two differential forms $dS = P^{mn} dg_{mn} - H_C dt$ and $dS = P^{\alpha\beta} dg_{\alpha\beta} - H_t dt$, 
where the dynamical variables are $\{ g_{\lambda\kappa}, P^{\alpha\beta} \}$, while the canonical $H_C$ and total $H_t$ Hamiltonians have been defined 
in the previous Section. These differential forms are included (as integrands) in the main integral invariant of mechanics, which is also known as the 
Poincar\'{e}-Cartan integral invariant \cite{Gant}. Now, by using these forms and procedure described in \cite{Gant} (see also \cite{Fro1}) one can 
derive (or restore) the original Jacobi equation. To simplify this (Jacobi) equation from the very beginning we introduce the new (local) temporal 
variable $d x_0 = \sqrt{- g} g^{00} dy_0$. In this variable the Jacobi equation takes the following form 
\begin{eqnarray}
 &-& \Bigl(\frac{\partial S}{\partial y_{0}}\Bigr) = I_{mnpq} \Bigl(\frac{\partial S}{\partial g_{mn}}\Bigr) 
 \Bigl(\frac{\partial S}{\partial g_{pq}}\Bigr) + \frac14 (- g) \Bigl[ I_{mnpq} B^{([mn] 0|\mu\nu k)} B^{(p q 0|\alpha \beta l)} - g^{00} 
 B^{\mu\nu k \alpha\beta l}\Bigr] g_{\mu\nu,k} g_{\alpha\beta,l} \nonumber \\ 
 &-& \frac12 \sqrt{-g} I_{mnpq} g^{mn} B^{(pq 0| \alpha\beta l)} g_{\alpha\beta,l} + \sqrt{-g} g^{00} T_2 \; , \; \label{eq5dHJ} 
\end{eqnarray}
where the explicit formula for the $T_2$ term is given by Eq. (\ref{eq5c}). This equation is the actual Jacobi equation (also called the Hamiltonian-Jacobi 
equation) for the free gravitational field in the metric GR, which has been derived in our new Hamiltonian formulation. As expected this Jacobi equation 
does not contain terms which are linear upon the partial $\frac{\partial S}{\partial g_{mn}}$ derivatives of the gravitational action $S$. 

As we have mentioned in \cite{Fro1} in analytical mechanics (see, e.g., \cite{Gant}, \cite{Arnold}) the methods based on the Jacobi equation are 
considered as the most effective procedures ever created to analyze the motion of an arbitrary, in principle, Hamiltonian system. It is also clear 
that all methods based on the Jacobi equation are usually very effective for dynamical systems with Hamiltonians which contain only a few relatively 
small powers of all essential momenta. This obviously includes the free gravitational field in metric GR, where the canonical and total Hamiltonians 
$H_C$ and $H_t$ are the quadratic functions of space-like momenta $\pi^{mn}$, Eq. (\ref{eq5}). The total Hamiltonian is also a linear function of 
$\sigma-$momenta $\pi^{0\sigma}$, or primary constraints $\phi^{0\sigma}$. To conclude this brief Section, let us note that if we draw an analogy with 
optics, then the momenta of the gravitational field ($\pi^{\alpha\beta}$, or $P^{\alpha\beta}$) should be called the tensor components of normal 
slowness tensor, while the gravitational action should be called the gravitational path length. In this language our Jacobi equation, 
Eq.(\ref{eq5dHJ}), plays the role of Huygens' principle for the free gravitational field. Although the meaning of similar analogies is quite limited.

\section{Discussions and Conclusion}

In this study we have developed the new, physically transparent and logically self-consistent 
Hamiltonian formulation of the metric gravity. In particular, we have have created an effective 
approach to determine various Poisson brackets which can now be used to perform a large amount 
of analytical and numerical calculations. The fundamental (or primary) Poisson brackets are 
defined between all components of the gravitational field and corresponding momenta (or 
components of the momentum tensor). The secondary Poisson brackets define commutation relations 
between arbitrary, in principle, analytical functions of coordinates (components of the 
gravitational field) and momenta. These Poisson brackets become the main working tools of the 
metric gravity, which can now be considered as an actual Hamiltonian system. Our Poisson 
brackets can be used to solve various problems in metric GR, e.g., obtain trajectories, derive 
and confirm new conservation laws, find integrals of motion, derive and investigate the laws of 
time-evolution for different quantities, vectors and tensors. 

Our approach allows one to determine the Poisson brackets from the two sets of basic dynamical 
variables: (a) set of straight dynamical variables of the metric gravity: $\{ g_{\alpha\beta}, 
\pi^{\gamma\rho} \}$ (or $\{ g_{\alpha\beta}, P^{\gamma\rho} \}$), and (b) dual set of basic 
dynamical variables of metric GR: $\{ g^{\alpha\beta}, \pi_{\gamma\rho} \}$ (or $\{ 
g^{\alpha\beta}, P_{\gamma\rho} \}$). We have found that these two sets of dynamical variables 
are always needed to construct the truly covariant and correct Hamiltonian formulations of the 
metric gravity. The fundamental relation between these two sets of dynamical variables is given 
by the Poisson bracket, Eq.(\ref{eq1551}). In our new dynamical variables the same relation 
takes the form $[ g_{\alpha\beta}, P^{\mu\nu}] = \Delta^{\mu\nu}_{\alpha\beta} = [ 
P_{\alpha\beta}, g^{\mu\nu}]$. The straight and dual sets of canonical (tensor) variables 
complement each other and they are crucially important to develop any non-contradictory
Hamiltonian approach to a system of interacting tensor fields, including the metric gravity. 

Another remarkable result obtained in this study should be emphasized here again: the canonical 
Hamiltonian $H_C$, which describes time-evolution of relativistic gravitational fields, can be 
reduced to its natural form, and this form is quadratic upon all essential momenta and coincides 
with the Hamiltonian of the non-relativistic system of $N (= d)$ interacting particles. Physical 
meaning of dynamical variables is obviously very different in both these cases, but almost 
identical forms of their Hamiltonians was absolutely unexpected. Briefly, the canonical 
Hamiltonian $H_C$ of the free gravitational field(s), Eq.(\ref{eq5a}), is reduced to the natural 
form, Eq.(\ref{eq5d}), which includes a pure quadratic function of the space-like momenta $P^{mn}$ 
with a positive coefficient in front of it. Indeed, the factor, which is located in front of the 
$P^{mn} P^{pq}$ product in the $H_C$ Hamiltonian, is the positively defined space-like tensor of 
the fourth rank $I_{mn pq}$ (or $\frac{1}{\sqrt{-g}} I_{mn pq}$). This factor can be considered as 
an effective inverse `quasi-mass' tensor of the free gravitational field in metric GR. Also, as 
directly follows from the explicit form of the canonical Hamiltonian $H_C$, Eq. (\ref{eq5d}), each 
of the remaining terms in this canonical Hamiltonian $H_C$ is a finite polynomial function of 
contravariant components $g^{\alpha\beta}$ of the metric tensor. The maximal power of such finite 
polynomials upon $g^{\alpha\beta}$ does not exceed eight. Some terms in the $H_C$ Hamiltonian also 
include the factors $\sqrt{-g}$ (or $\frac{1}{\sqrt{-g}})$ and $g^{00}$ (or $\frac{1}{g^{00}}$) 
component of the metric tensor.    

Note also that during our investigations we have constructed the set of new canonical $\{
g_{\alpha\beta}, P^{\gamma\rho} \}$ variables for the metric gravity. The total number of 
canonical variables equals $2 d$. The Poisson brackets between these variables are: $[ 
g_{\alpha\beta}, P^{\gamma\rho} ] = \Delta^{\gamma\rho}_{\alpha\beta} = \frac12 \Bigl( 
\delta^{\gamma}_{\alpha} \delta^{\rho}_{\beta} + \delta^{\rho}_{\alpha} 
\delta^{\gamma}_{\beta} \Bigr) = [ P_{\gamma\rho}, g^{\alpha\beta} ], [ g_{\alpha\beta}, 
g_{\gamma\sigma} ] = 0$ and $[ P^{\alpha\beta}, P_{\gamma\rho} ] = 0$. This indicates 
clearly that these new dynamical variables are truly canonical and can be used in the 
new Hamiltonian formulation of the metric gravity together with analogous set of dynamical 
variables $\{ g^{\alpha\beta}, P_{\gamma\rho} \}$ which is the dual set of canonical 
variables. 

To conclude our analysis let us formulate explicitly all essential, basic principles of the
Hamiltonian formulations of metric gravity, which must be fulfilled during construction of 
any working, physically significant and consistent Hamiltonian theory of the free 
gravitational field. For simplicity, these principles can be separated into three following
groups: (1) the general classical Hamilton-Jacobi principles generalized to the coupled tensor 
fields (see Section III above) which are applied to any Hamiltonian approach, (2) Dirac rules 
developed for dynamical systems with constraints, which includes the Dirac closure, and (3) 
conservation of the actual gauge symmetry during transition to the Hamiltonian approach (the 
Kiriushcheva-Kuzmin criterion of gauge conservation). Note that the principles from these 
three groups have extensively been discussed in this study. Nevertheless, here we repeat them 
to emphasize the crucial aspects of their definitions. First of all, we assume that there is 
the original non-singular Lagrangian which is written as an explicit function of all 
generalized coordinates and corresponding velocities. The first group of fundamental principles 
(or rules) tell us that transition from the original Lagrangian to the final Hamiltonian must 
be performed properly and unambiguously, e.g., with the use of the Legendre transformation. The 
arising Hamiltonian must be an explicit function of the momenta, which are conjugate to the 
corresponding velocities, and cannot include any of the essential velocities. Also, after 
transition to the final Hamiltonian(s) any introduction and/or injection of the new dynamical 
variables into these Hamiltonian is strictly prohibited, since it leads to fundamental mistakes 
(see, e.g., \cite{KK2011} and our Appendix B). 
 
These principles of Hamiltonian formulation must be applied in the natural order, i.e., from step 1 to step 3. This can be illustrated by the following 
example. Suppose you have introduced the new set of phase variables to describe the time-evolution of metric gravitational fields. The main question is
to show that your new momenta and coordinates are the canonical variables for the metric General relativity. At the first stage you need to check all 
Poisson brackets between all your variables in the both straight and dual phase spaces (see Section III above). If the new variables passed this step, 
then you need to proceed to the second step of the procedure and check the actual closure of the constraint chain. This means that the chain of arising 
constraints must be finite, i.e., after a number of steps all PB between constraints of highest order with the total Hamiltonian $H_t$ must explicitly be 
written as quasi-linear combinations of all constraints obtained at the previous steps of the procedure (Dirac closure). In addition to this, all PB 
between all essential constraints must be expressed as quasi-linear combinations \cite{QL} of these constraints and the total Hamiltonian $H_t$. If
this step has also been performed with no contradiction, then you are free to go to the last step of the procedure and check the conservation of gauge 
symmetry originally known for the given dynamical system with constraints. At this step by using all essential (primary, secondary, etc) constraints 
which have been derived at the previous step, one needs to show that the corresponding gauge generators, which are constructed with the help of Castellani 
procedure \cite{Cast}, allows one to restore the correct and complete gauge invariance (diffeomorphism) of the original system. In general, the last step 
is the most difficult step for actual checking, since all analytical computations here are very complex and require your constant and substantial 
attention. The diffeomorphism plays a central role in the Hamiltonian metric gravity, since here the components of metric tensor $g_{\alpha\beta}$ (or 
$g^{\alpha\beta}$) and corresponding momenta are used as the basic variables (not the physical coordinates $x^{\mu}$). The diffeomorphism guarantees that 
the real connection between the components of metric tensor $g_{\alpha\beta}$ (or $g^{\alpha\beta}$) and coordinates $x^{\mu}$ with each other is 
unambiguous and analytical (or smooth) at any time (see Appendix C). The only, but very substantial, indulgence is the fact that the checking of Dirac's 
closure and derivation of the diffeomorphism with the use of Castellani procedure can be performed on the zero-surface of primary constraints (on-shell). 

All known Hamiltonian formulations of the metric General Relativity are very complicated procedures, 
which are in dozens of times more complicated than analogous Hamiltonian formulations of the Maxwell 
electrodynamics. To operate successfully with the different Hamiltonian formulations of metric GR one 
needs to be familiar with the classical Hamiltonian procedures and tensor calculus. On the other hand, 
it is absolutely necessary to know well the both Dirac approach to the constrained dynamical systems 
and Castellani procedure which allows one to determine all generators of the actual gauge 
transformations. 

General principles of Hamiltonian formulation(s) are formulated here in the form which can be 
generalized to many other dynamical systems with constraints. For the metric gravity we have 
another crucial restriction, which follows from the fact that the structure and numbers of all 
essential, first-class constraints arising in the metric gravity are well known. Indeed, any 
new Hamiltonian formulation of the metric GR must lead to the $d$ primary constraints and $d$ 
secondary constraints \cite{Fro1} where $d$ is the dimension of our space-time manifold (time 
is always assumed to be scalar, one-dimensional variable). There is no way around this fact in 
the metric GR based on the $\Gamma - \Gamma$ Lagrangian, but the explicit forms of all these 
constraints in the new variables can be substantially different. As follows from our discussion 
of the fundamental principles of Hamiltonian formulation(s), currently, there are only 
three different, true canonical Hamiltonian formulations of the metric gravity: Dirac formulation 
\cite{Dir58} and Kiriushcheva-Kuzmin formulation \cite{K&K}. These two formulations are based on 
the use of rigorous dynamical variables (see Section V). Numerous quasi-Hamiltonian constructions 
created in this area of science since the end of 1950's are not canonical Hamiltonian formulations 
of the metric gravity. Therefore, it is useless to discuss that someone could `quantize' metric 
gravity by using similar quasi-Hamiltonian constructions. In addition to this, it is clear that 
quantization of the co- and contra-variant components of any tensor fields require a completely 
new procedure, since in quantum theory of metric gravity we have at least two different uncertainty 
relations. This follows form the fact that the two fundamental Poisson brackets $[ 
\pi^{\alpha\beta}, g_{\mu\nu}]$ and $[ \pi^{\alpha\beta}, g^{\mu\nu}]$ which must be obeyed 
simultaneously. There are some other similar facts which must be explained well before such a 
quantization of metric GR can be completed.       

Finally, as we all know many scientists called and considered the General Relativity (or metric GR 
in our words) as "the most beautiful of all existing physical theories" (see, e.g., \cite{LLTF}, 
page 228). Here we wish to note that the correct Hamiltonian formulation of the metric General 
Relativity (or, Gravity, for short) is also very beautiful physical theory. Furthermore, the truly 
covariant, very powerful and explicitly beautiful apparatus of this theory corrects everybody (even 
its authors), if they steps away from the unique, truly covariant and correct direction of actual 
theory. No comparison can be made with an ugly form of the original geometrodynamics (see, Appendix 
B) and other similar Hamiltonian-like creations, which were declared to be canonically related with 
the geometrodynamics. Note that Hawking in \cite{Hawk} called this `super-advanced' geometrodynamics 
by the theory which "contradicts to the whole spirit of General Relativity". As is shown in the 
Appendix B all these theories are based on the use of non-canonical variables. Therefore, all these 
theories and constructions have nothing to do with the actual Hamiltonian (metric) gravity. This 
explains the current catastrophe of Western gravity in application to many actual problems of metric 
General Relativity. Since 1959 more than 2000 papers were published in numerous journals about 
Hamiltonian formulations of the metric GR and their applications where their authors tried to predict 
and describe various gravitational phenomena. All these theories were based on the ADM variables, and 
on other similar sets of variables which are canonically related to ADM variables. This includes the 
so-called Ashtykar dynamical variables, variables used in Loop Quantum Gravity, etc (more details can 
be found in \cite{KK2011}). All these `advanced' variables are not canonical variables for the metric 
GR and cannot be transformed (canonically) in such variables. 

In addition to the use of non-canonical variables, ADM gravity and closely related theories (see, e.g., 
\cite{ADM} - \cite{Ish}) have a large number of troubling problems and spots \cite{KK2011}, e.g., the 
lost of complete diffeomorphism as a known gauge symmetry of the metric GR. Another closely related fact 
is: in ADM metric gravity one cannot apply all $d$ Bianchi identities for the Ricci tensor. These 
identities do obey in the original Einstein's metric gravity as well as in the Hamiltonian metric gravity 
developed by Dirac \cite{Dir58} and Kiriushcheva and Kuzmin \cite{K&K}, respectively. However, `somehow' 
in ADM metric gravity these Bianchi identities were lost (again we have to say this magic word `lost'!). 
The last fact is absolutely crucial to understand the general situation here: \textit{it is impossible to 
return from the ADM Hamiltonian for the free gravitational field to the original} $\Gamma - \Gamma$ 
\textit{Lagrangian by using only methods which are legally permitted to carry out such transitions} 
(more details can be found in \cite{Fro2022}). 

Finally, after 60 years of development and applications of `super-advanced' geometrodynamics and other 
similar `canonically equivalent' theories, we have to say that all these pure speculative `constructions' 
are incorrect and incomplete for solving actual problems of the metric gravity (for more details, see 
discussion in \cite{KK2011}). In addition tho this, the AMD metric gravity and other `super-advanced' 
theories are substantially different from the original Einstein's metric gravity. This is very 
embarrassing, since the correct Hamiltonian formulations of metric gravity developed in \cite{Dir58} and 
\cite{K&K} allow us to obtain the properties of the gravitational field which essentially coincide 
with the well known properties of such a field in the Einstein's metric gravity (No losses, no new 
findings).

I am grateful to my friends N. Kiriushcheva, S.V. Kuzmin and D.G.C. (Gerry) McKeon (all from the 
University of Western Ontario, London, Ontario, Canada) for helpful discussions and inspiration. \\

{\bf Appendix A} \\

In this Appendix we discuss relations between dynamical variables which are used in our and Dirac 
formulations of the metric General Relativity. In earlier paper \cite{FK&K} we have shown that 
dynamical variables $\{ g_{\lambda\kappa}, \pi^{\alpha\beta} \}$, which are used in the K$\&$K 
formulation of the metric GR, and analogous Dirac dynamical variables $\{ g_{\lambda\kappa}, 
p^{\alpha\beta} \}$ \cite{Dir58} are related to each other by some canonical transformation. That 
canonical transformation was written in the form \cite{FK&K} 
\begin{eqnarray}
 g_{\lambda\kappa} = g_{\lambda\kappa} \; \; \; {\rm and} \; \; \; p^{\alpha\beta} = 
 \pi^{\alpha\beta} - \frac12 \sqrt{- g} A^{(\alpha\beta) 0 \mu \nu k} g_{\mu\nu,k} 
 \; , \; \label{p_mom}
\end{eqnarray}
where the quantity $A^{(\alpha\beta) 0 \mu \nu k}$ is 
\begin{eqnarray}
  A^{(\alpha\beta) 0 \mu \nu k} = B^{((\alpha\beta) 0 \mid \mu \nu k)} - g^{0 k} 
  E^{(\alpha\beta) \mu\nu} + 2 g^{0 \mu} E^{(\alpha\beta) k\nu} \; , \; 
\end{eqnarray}
where $B^{((\alpha\beta) 0 \mid \mu \nu k)}$ is the $B^{(\alpha\beta 0 \mid \mu \nu k)}$ quantity 
(see, Eq.(\ref{Bcoef})) symmetrized in terms of all $\alpha \leftrightarrow \beta$ permutations.
Analogously, the $E^{(\alpha\beta) \mu\nu}$ and $E^{(\alpha\beta) k\nu}$ are the two symmetrized 
quantities (in respect to the $\alpha \leftrightarrow \beta$ permutations), i.e., 
\begin{eqnarray}
 E^{(\alpha\beta) \mu\nu} = e^{\alpha\beta} e^{\mu\nu} - \frac12 ( e^{\alpha\mu} e^{\beta\nu} 
 + e^{\alpha\nu} e^{\beta\mu} ) \; \; {\rm and} \; E^{(\alpha\beta) k\nu} = e^{\alpha\beta} 
 e^{k\nu} - \frac12 ( e^{\alpha k} e^{\beta\nu} + e^{\alpha\nu} e^{\beta k} ) \; , \; \nonumber
\end{eqnarray}
respectively, and $e^{\sigma\rho}$ are the Dirac tensors defined in Eq.(\ref{E}).  

As is shown in the main text the relation between our dynamical variables and dynamical K$\&$K 
variables introduced in \cite{K&K} takes the form $g_{\lambda\kappa} \rightarrow 
g_{\lambda\kappa}$ and $P^{\alpha\beta} \rightarrow \pi^{\alpha\beta}$, where 
\begin{eqnarray}
 P^{\alpha\beta} = \pi^{\alpha\beta} - \frac12 \sqrt{- g} B^{(\alpha\beta 0 \mid \mu \nu k)} 
 g_{\mu\nu,k} \; \; . \; 
\end{eqnarray}
From the last equation it is easy to obtain the following expression for our momenta 
$P^{\alpha\beta}$ written in terms of the Dirac momenta $p^{\alpha\beta}$  
\begin{eqnarray}
  P^{\alpha\beta} = p^{\alpha\beta} - \frac12 \sqrt{- g} \Bigl[ B^{([\alpha\beta] 0 
  \mid \mu \nu k)} - g^{0 k} E^{(\alpha\beta) \mu\nu} + 2 g^{0 \mu} E^{(\alpha\beta) k\nu} 
  \Bigr] \; , \; \label{P_p}
\end{eqnarray} 
where the quantity $B^{([\alpha\beta] 0 \mid \mu \nu k)}$ is the $B^{(\alpha\beta 0 \mid \mu 
\nu k)}$ coefficient, Eq. (\ref{Bcoef}), anti-symmetrized in respect to all permutations of 
the $\alpha$ and $\beta$ indexes. The transformation of dynamical variables $g_{\lambda\kappa} 
\rightarrow g_{\lambda\kappa}$ and $P^{\alpha\beta} \rightarrow p^{\alpha\beta}$, 
Eq.(\ref{P_p}), is the canonical transformation (this can be shown in the same way as it is 
done in the main text). Its inverse transformation is also canonical. This means that currently 
we have three different sets of dynamical variables which can be applied for the known and new 
Hamiltonian formulations of the metric GR: (a) Dirac variables \cite{Dir58}, (b) K$\&$K 
variables \cite{K&K}, and (c) our variables defined in Section VI of this study. These three 
different sets of dynamical variables are related to each other by simple canonical 
transformations. 

The canonicity of transformation of one set of dynamical variables into another such set is a 
necessary and sufficient condition for ordinary Hamiltonian systems. For Hamiltonian systems 
with constraints, this condition alone is no longer sufficient. An additional condition is 
formulated in the form that each constraint must be transferred into a similar constraint and 
vice versa, i.e., each primary constraint goes into the primary, while each secondary 
constraint goes into the secondary (for more detail, see, e.g., \cite{FK&K} and \cite{Fro1}). 
Currently, these two conditions should be considered as independent and completely sufficient 
for the transformation of dynamical variables to be canonical. This implies, in particular, 
that any correct Hamiltonian formulation of the metric Gravity must have $d-$primary and 
$d-$secondary constraints. Other Hamiltonian formulations of the metric General Relativity with 
different numbers of constraints, including formulations with tertiary constraints, are wrong 
$a$ $priori$. In addition to this, one can show that all possible Hamiltonian formulations with 
the even total number(s) of essential constraints are also wrong \cite{Fro1}. \\

{\bf Appendix B} \\

In this Appendix we want to show that dynamical variables which are used in geometrodynamics 
\cite{ADM} are not canonical. Therefore, this theory cannot be considered as the regular 
Hamiltonian formulation(s) of the metric GR. Furthermore, this theory, or geometrodynamics, 
cannot canonically be related to any of the correct Hamiltonian formulations currently known 
in the metric General Relativity. On the other hand, all similar `theories' which are 
canonically related to the geometrodynamics are equally wrong quasi-Hamiltonian constructions 
which cannot help anybody to solve problems arising in the metric General Relativity.  

The history of creation of geometrodynamics, which is also often called the ADM gravity, is 
straightforward. After an obvious success of Dirac paper \cite{Dir58} a small group of young 
authors, which included Arnowitt, Deser and Misner \cite{ADM}, decided to create some 
alternative (but Dirac-like!) formulation of the metric GR. Dynamical variables in this ADM 
approach were chosen as follows. The generalized six coordinates coincide with the 
corresponding space-space components $g_{pq}$ of the metric tensor $g_{\alpha\beta}$ defined 
in the four-dimensional space-time (or (3+1)-dimensional space-time, if we want to be 
historically precise). Four remaining coordinates were chosen in the form: the "lapse" $N = 
\frac{1}{\sqrt{- g^{00}}}$ and three "shifts" $N^{k} = - \frac{g^{0k}}{g^{00}}$, where $k = 
1, 2, 3$ (very likely, the idea to use these four coordinates was proposed by Wheeler). The 
corresponding space-like components of momenta $\Pi^{mn}$ were simply taken from Dirac paper 
\cite{Dir58} (see also our Appendix A), i.e., they coincide with the $p^{mn}$ momenta 
introduced by Dirac (see Appendix A). The four remaining momenta were not defined in the 
original ADM paper \cite{ADM}. Probably, this was done on purpose, since these four momenta 
lead to the primary constraints anyway, but ADM group has developed some special methods to 
operate with such `constraints' which included, in particular, the two important steps known 
as `constraints reshuffling' and `constraints solving' (like algebraic equations!). Right 
now, it is very hard to describe and discuss the internal logic of this quasi-theory, but we 
have to note that geometrodynamics was carefully analyzed earlier in \cite{KK2011} with many 
details and references. 

In fact, we do not need to dive into a deep discussion of ADM formulation, since we already 
have their ten generalized coordinates (one laps $N$, three shifts $N^{k}$ and six components 
of the metric tensor $g_{pq}$) and six momenta $\Pi^{mn}$ which coincide with the momenta 
$p^{mn}$ defined in Dirac's paper. By using only these dynamic variables of ADM gravity we 
can prove that these dynamical variables are not canonical. To prove this statement we have 
to calculate the two following Poisson brackets: (1) PB between "laps" $N$ and $\Pi^{mn}$ (or 
$p^{mn}$) momenta, and (2) PB between "shifts" and the same $\Pi^{mn}$ (or $p^{mn}$) momenta. 
If this theory is a truly Hamiltonian, then all these Poisson brackets must be equal zero 
identically. Let us check this simple fact. The first Poisson bracket is
\begin{eqnarray}
 &[&N, \Pi^{mn} ] = [ \frac{1}{\sqrt{- g^{00}}}, p^{mn} ] = - \frac{1}{\sqrt{(- 
 g^{00})^{3}}} [ g^{00}, p^{mn} ] \nonumber \\
 &=& 
 \frac{1}{\sqrt{(- g^{00})^{3}}} \; \; \frac12 ( g^{0 m} g^{0 n} +  g^{0 n} g^{0 m} 
 ) = \frac{1}{\sqrt{(- g^{00})^{3}}} g^{0 m} g^{0 n} \ne 0 \; , \; \label{N} 
\end{eqnarray}
while for the second bracket one finds
\begin{eqnarray}
 &[&N^{k}, \Pi^{mn} ] = [ -\frac{g^{0k}}{g^{00}}, p^{mn} ] = \frac{1}{2 g^{00}} 
 ( g^{0 m} g^{0 n} + g^{0 n} g^{0 m} ) - \frac{1}{( g^{00} )^{2}} g^{0 k} g^{0 m} 
 g^{0 n} \nonumber \\
 &=&  \frac{1}{2 ( g^{00} )^{2}} ( g^{0 0} g^{0 m} g^{k n} + g^{0 0} g^{0 n} g^{k m}
 - 2 g^{0 k} g^{0 m} g^{0 n} ) \ne 0 \; , \; \label{N-k} 
\end{eqnarray}
where $k = 1, 2, 3$. As follows from these equations none of these four Poisson brackets 
equal zero identically. Therefore, these dynamical variables are not canonical and theory 
which uses these variables is not a Hamiltonian theory of anything. Furthermore, it cannot 
be transformed into the correct Hamiltonian theory of metric GR, e.g., by applying some 
canonical transformation. Now, we can only guess that P.A.M. Dirac calculated the four 
Poisson brackets mentioned here in the end of 1950's (for him it took, probably, a few 
minutes). Even then Dirac could predict the sad failure of ADM approach (and other similar 
approaches) to the Hamiltonian formulation of metric GR in the future. Now that future has 
become our present. \\

{\bf Appendix C} \\

In this Appendix we discuss some important technical details which are crucially important 
to construct different Hamiltonian approaches to the metric gravity, but we could not 
include them in the main text. First, by using some help from Lanczos \cite{Lanc} we develop 
the Hamilton approach for those cases when we have to operate with variations of the original 
Lagrangian only and cannot use (for some reasons) the Lagrangian function itself (see, e.g., 
Section II in the main text). To achieve this goal we represent the variation of the original 
Lagrangian $\delta L$ in the form 
\begin{eqnarray}
    \delta L = \delta ( v_i p_i - H ) = p_i \delta v_i + v_i \delta p_i - \delta H \; , 
    \; \label{ApCf1}
\end{eqnarray}
where $L$ is the function of the generalized coordinates $q_i$ and velocities $v_i = 
\frac{d q_i}{dt}$ only, while $H$ is the function of the same coordinates $q_i$ and new 
variables (momenta) $p_i$ only. This means that: \\
(1) an arbitrary variation of the momentum $p_i$ has no influence on the variation of $L$. \\
(2) an arbitrary variation of the velocity $v_i$ has no influence on the variation of $H$. \\ 
At this point all momenta must be considered as some unidentified functions. 

First, we re-write Eq.(\ref{ApCf1}) to the form 
\begin{eqnarray}
    \delta L - p_i \delta v_i = v_i \delta p_i - \delta H \; , \; \label{ApCf1a}
\end{eqnarray}
or 
\begin{eqnarray}
  \frac{\partial L}{\partial q_i} \delta q_i + \frac{\partial L}{\partial v_i} \delta 
  v_i  - p_i \delta v_i = v_i \delta p_i - \frac{\partial H}{\partial q_i} \delta q_i 
  - \frac{\partial H}{\partial p_i} \delta p_i  \; . \; \label{ApCf1b}
\end{eqnarray}
From here one finds 
\begin{eqnarray}
  \frac{\partial L}{\partial q_i} \delta q_i + \Bigr( \frac{\partial L}{\partial v_i} 
  - p_i \Bigr) \delta v_i =    \Bigl( v_i - \frac{\partial H}{\partial p_i} \Bigr) 
  \delta p_i - \frac{\partial H}{\partial q_i} \delta q_i \; . \; \label{ApCf2}
\end{eqnarray}
Now, the point (1) leads to the equation $p_i = \frac{\partial L}{\partial v_i}$ which is, 
in fact, the definition of momenta. Analogously, based on the point (2) one finds another 
equation $v_i = \frac{\partial H}{\partial p_i}$ which is the definition of the velocities 
in the phase space. After these steps we also obtain the third equation $\frac{\partial 
L}{\partial q_i} = - \frac{\partial H}{\partial q_i}$. If the both functions $L$ and 
$H$ explicitly depend upon time $t$ and an additional parameter $\alpha$, then here we can 
write analogous equations $\frac{\partial L}{\partial t} = - \frac{\partial H}{\partial t}$ 
and $\frac{\partial L}{\partial \alpha} = - \frac{\partial H}{\partial \alpha}$. These three 
groups of equations are the basic equations of the Hamilton procedure. To make this system 
of basic equations a complete system of dynamical equation one has to add the system of 
Lagrange equations of actual motion $\frac{d}{dt} \frac{\partial L}{\partial v_i} = 
\frac{\partial L}{\partial q_i}$ which in Hamiltonian variables is written in the form 
$\frac{d p_i}{dt} = - \frac{\partial H}{\partial q_i}$. Finally, we can write all actual 
dynamical equations in the Hamilton method $\frac{d q_i}{dt} = \frac{\partial H}{\partial 
p_i}$  $\frac{d p_i}{dt} = - \frac{\partial H}{\partial q_i}$. These equations are logically 
closed and contain neither the original Lagrangian $L$, nor the velocities $v_i$. Now, we 
can call and consider the function $H$, which appears in these equations, as the Hamiltonian 
of the system. 

Another crucial thing which we need to discuss here is the chain and multi-chain integrals 
over different components of the metric tensor. First, consider the following one-dimensional 
metric integral 
\begin{eqnarray}
 \int F^{\alpha\beta} dg_{\beta\gamma} = \int F^{\alpha\beta} \frac{\partial 
 g_{\beta\gamma}}{\partial x^{\mu}} dx^{\mu} = \int F^{\alpha\beta} \Bigl( 
 \Gamma_{\gamma,\beta\mu} +  \Gamma_{\beta,\gamma\mu} \Bigr) dx^{\mu} \; \; \label{ApCf3}
\end{eqnarray}   
where $\Gamma_{\gamma,\beta\mu}$ are the Cristoffel symbols of the first kind, while the last 
integral in the right-hand side is the usual (linear) integral taken in the $d-$dimensional 
coordinate space. This formula can be generalized to the two-, three- and multi-dimensional 
metric integrals. For instance, in the two-dimensional case we have to replace 
$dg_{\beta\gamma} dg_{\lambda\sigma} = \Bigl( \Gamma_{\gamma,\beta\mu} + 
\Gamma_{\beta,\gamma\mu} \Bigr) \Bigl( \Gamma_{\sigma,\lambda\nu} + 
\Gamma_{\lambda,\sigma\nu} \Bigr) dx^{\mu} dx^{\nu}$. 

However, it is easy to note that the both sides of Eq.(\ref{ApCf3}) are transformed differently 
during general transformations of the $x^{\alpha}$ coordinates. Indeed, in the left-hand side 
of this equation we have an absolute (or first-class) variable $g_{\beta\gamma}$, while in the 
right-hand side of the same equation we have a set of usual (or second-class) variables 
$x^{\mu}$, where $\mu = 0, 1, \ldots, d - 1$. In general, by varying these $x^{\mu}$ coordinates 
arbitrarily one quickly arrives to fundamental contradictions with the use of Eq.(\ref{ApCf3}). 
To avoid such situations and define all multi-chain integrals over different components of the 
metric tensor correctly and uniformly we need to introduce some special sets of $x^{\mu}$ 
variables (or coordinates) which are called the Killing's sets, or curves. The Killing's curves 
consist only those $x^{\mu}$ variables which do not change metric, i.e. all components of metric 
tensor, under infinitesimal transformations. For general transformations of variables in 
metric GR which are written in the form $x^{\mu} \rightarrow x^{\mu} + \xi^{\mu}$, where 
$\xi^{\mu}$ are the small values, the Killing's criterion is written in the form $\nabla^{\mu} 
\xi^{\nu} + \nabla^{\nu} \xi^{\mu} = 0$, where $\nabla^{\mu} \xi^{\nu} + \nabla^{\nu} \xi^{\mu}$ 
is the sum of contravariant derivatives of the $\xi^{\nu}$ and $\xi^{\mu}$ values (for more 
details, see \cite{LLTF}). In turn, the Killing's criterion is equivalent to the diffeomorphism 
gauge symmetry (or diffeomorphism conservation) in the metric GR \cite{K&K}. Thus, the 
diffeomorphism plays a central role for the both Hamiltonian formulation(s) of the metric GR and 
for correct and uniform definition of the chain integrals over components of the metric tensor. 
The central role of diffeomorphism in the metric gravity follows from the fact that the variables 
in metric gravity are the components of metric tensor $g_{\alpha\beta}$ (or $g^{\alpha\beta}$), 
but not the physical coordinates $x^{\mu}$. The diffeomorphism guarantees that the real 
connection between the components of metric tensor $g_{\alpha\beta}$ (or $g^{\alpha\beta}$) and 
coordinates $x^{\mu}$ with each other is unambiguous and analytical (or smooth) at any time.


\begin{thebibliography}{99}
\bibitem{QL} The coefficients in such `quasi-linear' combinations are field-dependent, i.e., they 
are neither numbers, nor `group' constants.
\bibitem{Hamlt} There is an obvious logical gap here, since we have replaced the actual Einstein-Hilbert 
Lagrangian by its variational equivalent, i.e., by the $\Gamma - \Gamma$ Lagrangian. Briefly, this means 
that after such a replacement we have to operate with the variations of Lagrangian(s) only. The Hamilton 
approach for variations of the original Lagrangian is developed in the Appendix C.
\bibitem{canon} The word `canonical' originated from the Greek word $\kappa \alpha \nu o \nu \iota 
\kappa o \varsigma$ which means correct, equal, equivalent and true.
\bibitem{constr} It can be shown that all primary $\phi^{0\lambda}$ and secondary constraints 
$\chi^{0\sigma}$ which arise during this Hamiltonian procedure are the first class constraints 
\cite{Dir50}.
\end{thebibliography}
\end{document}